\documentclass[pre,amssymb,amsmath,twocolumn,eqsecnum,floatfix,showpacs]{revtex4}
\usepackage[thinspace,thinqspace,textstyle,pstricks]{SIunits}
\usepackage{graphicx,psfrag}
\usepackage{color}
\usepackage{pstricks,pst-node}

\begin{document}

\newcommand{\cum}[1]{\ensuremath\langle\langle #1\rangle\rangle}
\newcommand{\ham}{\ensuremath\mathcal{H}}

\newcommand{\str}{\ensuremath_\mathrm{stretch}}
\newcommand{\bend}{\ensuremath_\mathrm{bend}}
\newcommand{\bind}{\ensuremath_\mathrm{bind}}
\newcommand{\steric}{\ensuremath_\mathrm{steric}}
\newcommand{\el}{\ensuremath_\mathrm{el}}

\newcommand{\Nbar}{{\bar N}}
\newcommand{\Rbar}{{\bar R}}
\newcommand{\Pbar}{{\bar P}}

\newcommand{\uvec}[1]{\hat{\boldsymbol{#1}}}
\renewcommand{\vec}[1]{\boldsymbol{#1}}
\newcommand{\eps}{\ensuremath\epsilon}

\newcommand{\vkl}{\ensuremath\ell_f}
\newcommand{\scl}{\ensuremath\ell_\theta}
\newcommand{\Ybulk}{\ensuremath Y} 
\newcommand{\kappabulk}{\ensuremath\kappa}
\newcommand{\Ydisc}{\ensuremath{\tilde Y}}
\newcommand{\kappadisc}{\ensuremath{\tilde \kappa}}
\newcommand{\vK}{\ensuremath \gamma}

\newcommand{\SH}[1]{\textit{\color{blue}[#1]}}
\newcommand{\todo}[1]{{\bf\color{green}[#1]}}
\newcommand{\later}[1]{\textit{[#1]}}
\newcommand{\CLH}[1]{\textit{\color{red}[#1]}}

\newcommand{\fd}{{\it fd}}

\renewcommand{\SH}[1]{}
\renewcommand{\later}[1]{}
\renewcommand{\CLH}[1]{}

\newcounter{fnnum}
\setcounter{fnnum}{1}
\renewcommand{\footnote}[1]{\expandafter\def\csname footnote\Alph{fnnum}\endcsname{#1}\cite{footnote\Alph{fnnum}}\addtocounter{fnnum}{1}}
\newcommand{\citefootnote}[2]{\expandafter\def\csname footnote\Alph{fnnum}\endcsname{#2}\cite{#1,footnote\Alph{fnnum}}\addtocounter{fnnum}{1}}

\title{An irreversible growth model for virus capsid assembly}
\author{Stephen D. Hicks}
\email{sdh33@cornell.edu}
\author{C. L. Henley}
\email{clh@ccmr.cornell.edu}
\affiliation{Department of Physics,
  Cornell University,
  Ithaca, New York 14853}

\begin{abstract}
We model the spontaneous assembly of a capsid (a virus' closed outer
shell) from many copies of identical units, using entirely
irreversible steps and only information local to the growing edge.
Our model is formulated in terms of (i) an elastic Hamiltonian with
stretching and bending stiffness and a spontaneous curvature, and (ii)
a set of rate constants for addition of new units or bonds.  An
ensemble of highly irregular capsids is generated, unlike the
well-known icosahedrally symmetric viruses, but (we argue) plausible
as a way to model the irregular capsids of retroviruses such as HIV.
We found that (i) the probability of successful capsid completion
decays exponentially with capsid size; (ii) capsid size depends
strongly on spontaneous curvature and weakly on the ratio of the bending
and stretching elastic stiffnesses of the shell; (iii) the degree of
localization of Gaussian curvature (a measure of facetedness) depends
heavily on the ratio of elastic stiffnesses.
\end{abstract}


\maketitle

\section{Introduction}
In recent years the question of spontaneous assembly has arisen in
many apparently disconnected fields, including
nanofabrication~\cite{kim03, cheng04}, robotics and
microelectronics~\cite{gracias00,whitesides02}, and particularly
biology~\cite{kushner69,perham75}.
%Among biological systems that
%assemble themselves \SH{maybe more specific citations?}  are protein
%molecules, microtubules, and actin.
%For many biological structures, assembly requires energy in the form
%of ATP hydrolysis~\cite{carlier90,hartl95,ito97}, but for numerous
%others it runs spontaneously: in particular that of lipid
%bilayers~\cite{kimizuka01,israelachvili77} and of virus
%capsids~\cite{salunke86,bruinsma03}, the subject of this paper.
While the assembly of many biological structures, such as actin
filaments and chromatin, requires energy in the form of ATP
hydrolysis~\cite{carlier90,hartl95,ito97}, numerous other 
structures assemble spontaneously.  In particular are lipid
bilayers~\cite{kimizuka01,israelachvili77} and virus
capsids~\cite{salunke86,bruinsma03}, the subject of this paper.

\subsection{Quasiequivalence}
\label{sec:intro-quasi}
The capsid of a virus is the shell of proteins surrounding and
protecting the viral genome (DNA and RNA).  Capsids are observed in a
wide variety of sizes, ranging into the thousands of proteins; most
known capsids have icosahedral or cylindical point group
symmetry~\citefootnote{reddy98}{It was noted early that capsids ought
to be built from many identical copies of comparatively small
proteins, in order to maximize the volume available for the genome,
while minimizing the space on the genome needed to code
them~\cite{crick56}.  However, point group symmetries set an upper
limit of 60 units which can be joined equivalently to form a closed
convex polyhedron~\cite{crick56,caspar62}; thus many proteins must be
inequivalent by the symmetry}.  A typical virus uses only one, or a
few, kinds of protein in its capsid; consequently noted, typical
capsids are necessarily built from copies of the same unit in
positions that are \emph{not} equivalent by any global
symmetry~\cite{caspar62}.  However, \citet{caspar62} identified an
elegant approximate symmetry which they called
\emph{quasiequivalence}.  The key idea is that locally, every bit of
the capsid is a patch of triangular lattice; in an infinite triangular
lattice, all the units \emph{would} be symmetry
equivalent.  They argued that typical
proteins could accomodate a variation of $\pm 5\degree$ in bond
angles~\citep{pauling53}, while maintaining the same microscropic
bonding between proteins.  This allows representation of any capsid as
a network of approximately equilateral triangles, with a constraint
(due to the bond angle limitation) that the number of triangles around
every vertex must always be either five or six.  The points of local
five-fold symmetry may be identified with the topological defects
called \emph{disclinations} (to be defined in
Sec.~\ref{sec:model-conf}), and any closed shell must contain exactly
twelve of them.  In an icosahedral capsid, the disclinations form the
vertices of a large icosahedron, the edges of which have length
$\sqrt{T}$ in lattice units, where the triangulation number $T=1,3,4,
7,\ldots$ is one of a sequence of discrete allowed
integers~\cite{caspar62}, so that there are $60T$ small triangles.

We emphasize that the rules of quasiequivalence do not 
force any global symmetry, nor do they fix the size of
the completed capsid.
Thus it is surprising that many viruses reliably assemble
large symmetric capsids. The challenge to theory is to 
explain both the size and shape selection, or at
least to explain why a closed shell is formed, when 
tubes or sheets would be equally consistent with the
local bonding.
It would not be surprising if models predict different 
capsids depending on parameters (which might experimentally
correspond to pH, salt content, catalysts, protein
concentrations, or mutations in the capsid protein).
Such polymorphic behavior is very fruitful to study in quasiequivalent
models: it effectively explores more of the 
various local geometries in which the proteins can bind
and thus can allow more parameters to be determined, in principle.
This paper develops a model of irreversible (non-equilibrium) assembly of 
quasiequivalent units which produces a highly polymorphic
ensemble of capsids, which we argue below may model the growth
of retrovirus capsids.

It is worth mentioning that \citet{twarock04} has developed virus
tiling theory, an extension to Caspar--Klug quasiequivalence which 
uses rhombs and kites rather than triangles, and can therefore describe
the anomalous viruses from \emph{Papovaviridae}.

\subsection{Recent Models}
\label{sec:intro-models}
The most successful recent capsid models consistent with
quasiequivalence are \emph{equilibrium} theories: a microscopically
motivated phenomenological Hamiltonian is shown to be optimized by
certain shapes, and it is assumed that this free energy minimum is
found during the actual assembly process.  Thus Bruinsma et~al~\cite{bruinsma03,zandi04} modeled pentamers and hexamers as
different-sized discs packed on a sphere, with an effective
Hamiltonian favoring dense packing, a bending stiffness with
spontaneous curvature, and a switching cost to make pentamers (rather
than hexamers) of the proteins.  When this switching cost is small,
icosahedral viruses were selected over nonicosahedral
shapes~\cite{zandi04}.  Additionally they demonstrated polymorphism,
similar to phenomena seen in Cowpea Chlorotic Mottle Virus (CCMV), by
showing a phase transition between tubes, $T=3$, and $T=1$ capsids as
the model parameters varied.  Another family of models, introduced by
Nelson \cite{lidmar03}, focuses on the external shape of large capsids, using
continuum elastic theory: the shape evolves from practically spherical
to sharply faceted as the size increases or the bending stiffness
decreases.

Alternate theories to quasiequivalence have also been developed,
still in terms of an equilibrium picture.  Most
notable is the ``local rules'' theory~\cite{berger94,schwartz98,schwartz00}
which posits \emph{several} ``flavors''
of unit, with \emph{inequivalent} edges,
and  rules for the joining of these different kinds of edges
so the units fit together like puzzle pieces
and there is a single unique structure that 
obeys all the matching rules.  
It is generally necessary to assume that 
the same capsid protein molecule has different conformation
species, each of which has entirely different specific binding.  
It appears implausible that so many different functions
could be built into one molecule, or that evolution
could have discovered this solution, if it is the only
way to engineer a large capsid.
We therefore prefer a theory without such matching rules.

Another class of theory focuses on the \emph{process} of 
assembly, which one might expect is far from equilibrium.
In particular, \citet{zlotnick94,zlotnick05} has focused on the kinetics 
of capsid growth.
Using a basic unit of $5T$ proteins, so that the complete capsid is a
dodecahedron of 12 units, and a free energy based on the number of
adhered edges, he considers the species of the most stable incomplete
capsid of any size and constructs rate equations relating the
concentrations of each species.  This leads to the phenomenon of the
kinetic trap: if the initial concentration of monomers is too large,
they aggregate quickly into larger structures, slowing the later 
stages of growth since the required monomers are depleted.

Similar kinetic models have been extended using virus tiling
theory\cite{twarock04}.  \Citet{keef05} extend Zlotnick's work
and consider the effect of different association energies on the 
kinetics of \emph{Papovaviridae} assembly.

So far we have limited our discussion to icosahedral viruses.  While
there is some polymorphism in icosahedral viruses -- usually changing
$T$ numbers under different conditions -- the capsids are still
generally symmetric.  Mature retroviral capsids, on the other hand,
have been observed to be very irregular~\cite{ganser99}.

\subsection{Retroviruses}
\label{sec:intro-retro}
Retroviruses, such as human immunodeficiency virus (HIV) and rous
sarcoma virus (RSV), are RNA viruses which all contain a
characteristic enzyme -- reverse transcriptase -- allowing the RNA to be
transcribed into DNA for infection.  Upon infection, the virus
produces copies of several proteins -- in particular, the structural
polyprotein Gag.  Many copies of Gag (approximately 1500 for 
RSV~\cite{vogt99} or 5000 for HIV~\cite{briggs04}) aggregate at the cell
membrane before budding out of the cell as an immature virus particle.
A maturation step then takes place in which a protease cleaves Gag
into its constituent proteins: matrix (MA), capsid (CA), and
nucleocapsid (NC).  After cleavage, the MA remain bound to the lipid
membrane and the NC remain bound to the RNA.  A fraction of the CA
then reassemble into the mature capsid~\cite{briggs03}.  In HIV, this
fraction has been measured to be roughly 30\%~\cite{briggs04,lanman04}.

\Citet{ganser99} proposed a
model to point out that the cones formed by mature HIV cores must have
quantized angles and measured this on electron micrographs.  Later
studies have measured these angles and other data using more accurate
tomography~\cite{briggs06,benjamin05}.  In addition to the irregular
structure, polymorphism is also observed in the switching between
tubes, cones, and spheres under different conditions~\cite{ehrlich01}.
RSV, on the other hand, has cores that are observed to be roughly
spherical, but with a wide distribution in the degree of
asphericity~\cite{kingston01}.

There is significant variation in size and shape of mature HIV
capsids, but they are commonly cones.  Following the quasiequivalence
paradigm, these are described geometrically by locally triangular
lattices like carbon fullerene cones~\cite{ganser04,ge94}.  There is
no well-established explanation of the cone's shape or size.
\Citet{briggs06} suggest that the small end of the cone forms first,
possibly from a sort of template, and that the large end forms when
the growing capsid runs into the membrane.  \Citet{benjamin05} instead
suggest that the large end is nucleated first.  \Citet{nguyen05}
developed an equilibrium theory combining the ideas of
Refs.~\cite{bruinsma03} and~\cite{lidmar03}, adding fullerene
cones~\cite{ge94} to consideration.  Assuming a fixed size, they
generate a family of configurations of maximum symmetry, and find a
phase transition between cones, tubes, and spheres as a function of
the elastic parameters.  A weakness of their model, however, is that
cones are stable in a relatively small portion of parameter space, and
their appearance at all depends critically on the assumption of
fixed-size, which is unphysical.  In HIV maturation, only a third of
the CA proteins assemble into the mature conical capsid, leaving the
rest in solution within the virus' lipid envelope
\cite{briggs03,briggs04}.  Thus, we should expect the capsid size to
vary freely.  Later work by \citet{nguyen06} suggests that conical
capsids are not energy minima, but are instead selected by assembly
constraints.

\subsection{Outline}
In the following pages we present our model of capsid assembly, discussing
our choice of energy and transitions which govern growth of a capsid from
a single unit to a complete closed shell, by alternately minimizing the
energy and choosing a transition to a larger capsid.

We discuss common failure modes in this model and the choices of
parameters in which they arise.  In particular, we look at a mostly
avoidable failure which occurs at the end of growth in which a small
hole cannot be completed, and a more problematic failure which occurs
in the middle stages of growth and involves narrow ``fingers'' of
capsid.

We then consider a number of ways to measure growing and complete
capsids, largely motivated by experimental measurements.  We present
three main results.  First, capsid size depends primarily
on the spontaneous curvature, but also on the ratio of elastic
constants.  Second, growth failure is a rougly Poissonian
process, and thus the probability of successful growth decreases
exponentially with the expected size.  Third, we discuss the
application of our model to various measurements of the shape of
capsids, with particular emphasis on the Gaussian curvature.

Finally, we summarize our results and discuss the advantages and
disadvantages of our model, and possible future directions.

\section{Irreversible Growth Model}
\label{sec:model}
We now introduce a model to describe quasiequivalent capsid assembly
in a far-from-equilibrium picture.  Consider a single growing capsid
and a number of units in solution.  We picture the units slowly
accreting onto the growing capsid until the finished product is formed.

Our choice is to represent this by adopting the simplest possible
model that can represent a growing capsid and be simulated
efficiently: this precludes representing each protein as a rigid body
moving in space.  Instead, a capsid (growing or completed) is
represented by a triangular network (Sec.~\ref{sec:model-conf}),
with an elastic energy governing the bond lengths and angles
(Sec.~\ref{sec:model-ham}).  We do not explicitly represent the
units in solution, instead formulating a set of first-order rate
equations for addition of a unit to the capsid or for other discrete
changes in the network geometery (Sec.~\ref{sec:model-growth}).

Other physical or mathematical models have been abstracted to a
similar degree~\cite{bruinsma03,zandi04,lidmar03}, following a
standard philosophy of statistical mechanics.  Some capacity to adapt
the model to (say) a specific virus species is lost, but the
simplicity makes it conceivable to grasp the physical meaning of each
parameter, and feasible to explore all dimensions of the parameter
space by simulations.  Typically, only particular combinations of the
microscopic parameters matter, and a properly formulated toy model
adopts those combinations.  It can happen that fairly different
microscopic systems may, through such elimination of unimportant
parameters, all map to the same simple model; in that case, the model
offers a possibility of unifying the description of all these systems.

\subsection{Configuration degrees of freedom}
\label{sec:model-conf}

Our formulation depends on {\it two} complementary kinds of degree of
freedom, a discrete kind we call ``topological'' and a continuous kind
called ``positional''.  The former consists of a bond network built
from triangles, with vertices either connected by a bond or not; the
latter consists of the actual coordinates of the vertices in space.
Since prior work emphasized equilibrium, we took the opposite limit by
allowing no change in any bond, once formed.  One consequence is that
our discrete ``topological'' variables are more fundamental than the
positional ones: given a network of bonds, the angles and bond lengths
will relax to a constrained minimum (or fluctuate thermally around it)
as determined by a Hamiltonian, defined in Sec. \ref{sec:model-ham}.
In our growth model, these positional variables feed back into the
discrete ones by controlling the relative rates of alternative changes
in the network as units are accreted.  (In principle one could
envisage a further abstraction in which the positional variables are
eliminated completely and the rates expressed directly in terms
of the bond topology, but we did not attempt that.)

The models discussed above in Sec. \ref{sec:intro-models} all have
essentially just a single type of degree of freedom -- the first, the
second, or something intermediate.  \Citet{lidmar03} assume a
predetermined graph topology, so only the vertices' positions are
nontrivial.  On the other hand, \Citet{endres05} discard position
information and only consider the (discrete) connectivity.  
%\CLH{I
%consider Bruinsma to be sort of intermediate or mixed.  In particular,
%his penton energy is an attribute of the discrete geometry.  In
%effect, their variables are discrete up to the pentamer/hexamer
%radius, and continuous from there out -- ???}  
Finally, \citet{bruinsma03}
continuously vary the positions of the discs, and determine which
discs neighbor one another secondarily.

\begin{figure}
  \includegraphics[width=0.8\columnwidth] {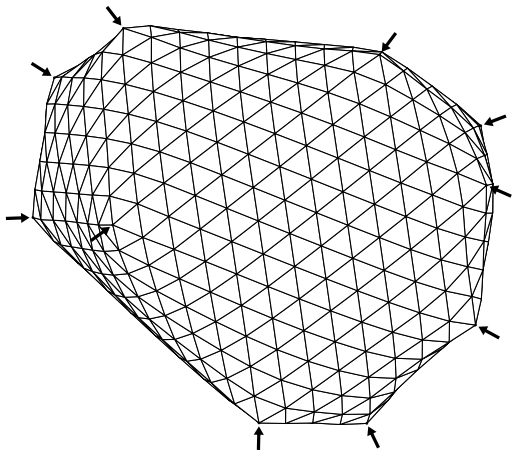}
  \caption{\label{fig:capsid}
    Example of a closed final capsid resulting from our 
    growth simulation, with $\vkl=.12r_0$, $\theta_0=16\degree$,
    $\Gamma_{I,J}=50$, and $\sigma=12.7\degree$.  These parameters
    are explained in Secs.~\ref{sec:model-ham}-\ref{sec:model-growth}.
    The disclinations are marked by arrows.}
\end{figure}

It may be questioned why we have chosen triangles as the fundamental
building blocks.  In a model more faithful to a particular virus
species, one would want to add the multimer which is accreted in
nature.  Virus species assembling from
dimers~\cite{zlotnick00,speir95}, trimers~\cite{forrer04}, and
pentamers/hexamers~\cite{xie95}, have all been observed
experimentally.  HIV has so far formed only dimers in
solution~\cite{vonschwedler03}.  Several groups have done molecular
dynamics simulations using solutions of
monomers~\cite{berger94,rapaport04} and kinetic simulations with
pentamers~\cite{zlotnick94}, dimers~\cite{endres02}, or
trimers~\cite{zlotnick03a}.  Because of its simplicity, and the work
done on tethered surfaces by Nelson and
coworkers~\cite{seung88,lidmar03}, we will focus on a trimer-based
model for this initial work, an example of which can be seen in
FIG.~\ref{fig:capsid}.  It is at this point worth noting that the
hexagonal lattice which is dual to our triangular lattice is in fact
very similar to the molecular lattices formed from HIV CA~\cite{li00}.

%Our use of triangular units, rather than e.g. trapzoids (to
%represent monomers~\cite{rapaport04}) or hexamers/pentamers
%(as in Refs.~\cite{bruinsma03} and \cite{zandi04}) is also
Our use of triangle units is also influenced by the
notion of ``universality'' in physics, whereby the functional form of
elastic theory, or the critical exponents of a phase transition, are
independent of the particular lattice used at the microscopic scale.
In any of the alternative representations, one can still define a
triangular, locally sixfold lattice with rare locally 5-fold points in
it. Much experience in statistical mechanics suggests that, at
``coarse-grained'' length scales (those large compared to the lattice
spacing), the behavior stops depending on the details.  However, two
related caveats must be expressed, that (i) possibly a detail of the
microscopic model forces a certain parameter of the coarse-grained
model to be strictly zero, thereby changing the qualitative behavior
(``universality class''); (ii) it may be that a parameter regime easy
to achieve in one version of the model will require a complicated
fine-tuning of parameters in an alternate version.

\Citet{caspar62} noted that only pentamers and hexamers have small
enough deformations to be allowed in quasiequivalence, and thus any
quasiequivalent capsid must have exactly twelve pentamers.
Quasiequivalence is based on a flat triangular lattice, so that a
pentamer is a {\it disclination}: a topological defect of the
triangular lattice. This means it can be characterized by effects at
an arbitrary distance; namely, if we parallel transport a vector
around a loop, that vector ends up rotated by $(\pi/3) N_{\rm disc}$
from its starting orientation, where $N_{\rm disc}$ is the number of
5-fold disclinations enclosed by the loop.  (This is called a
``disclination charge'' by analogy to how the electric charge enclosed
by a surface is determined by an integral of the electric field over
that surface, according to Gauss's law.)  That there are exactly
twelve disclinations can now be seen either by counting vertices,
edges, and triangles under the constraint $V-E+F=2$, or more generally
because the total disclination charge must sum to
$4\pi$~\footnote{These are both consequences of the Euler
characteristic for a genus~0 surface, $\chi(g)=2-2g=2$.  More
generally, $V-E+F=\chi(g)$, and the total disclination charge must sum
to $2\pi\chi(g)$.  For more information, see Eric W. Weisstein. ``Euler
Characteristic.'' From MathWorld--A Wolfram Web
Resource. http://mathworld.wolfram.com/EulerCharacteristic.html
and references therein}.

Between the disclinations are patches of regular sixfold 
lattice with no topological freedom:  thus, 
\emph{the capsid is completely determined by the placement of the
disclinations}.  Since there may be hundreds of network 
vertices, and only twelve disclinations, this is in principle
a simplification.

\subsection{Hamiltonian}
\label{sec:model-ham}

We represent the growing capsid as a number of approximately
equilateral triangles connected along the edges.
We then generalize the discretized Hamiltonian used by
\citet{lidmar03} to include spontaneous curvature $\theta_0$ and
steric terms.
\begin{equation}
  \ham = \ham\str + \ham\bend + \ham\steric .
\label{eq:ham}
\end{equation}

\subsubsection{Elastic energy}
The first two terms in eq.~(\ref{eq:ham}) are elastic
terms for bond stretching and bending:
\begin{equation}
  \ham\str = \frac{\sqrt{3} \Ydisc}{4}\sum_{\langle ij\rangle}
    \left(\left|\vec r_i-\vec r_j\right| - r_0\right)^2, 
\label{eq:ham-str}
\end{equation}
\begin{equation}
  \ham\bend = \frac{2 \kappadisc}{\sqrt{3}} \sum_{\langle IJ\rangle}
    \left(1-\cos\left(\theta_{IJ}-\theta_0\right)\right) .
\label{eq:ham-bend}
\end{equation}
Here, $\langle ij\rangle$ denote pairs of nearest-neighbor
vertices with positions $\vec r_i$, and $\langle IJ\rangle$ denote
pairs of nearest-neighbor triangles.  The exterior dihedral angle
\begin{equation}
  \theta_{IJ}=\cos^{-1}\left(\uvec n_I\cdot\uvec n_J\right),
\end{equation}
 where $\uvec n_I$ is the unit normal to triangle $I$.

Our discretized parameters, $\Ydisc$ and $\kappadisc$, have the same
dimensions as the two-dimensional Young's modulus $\Ybulk$, and
bending stiffness $\kappabulk$, respectively, which are emphasized in
continuum approaches to predicting capsid shapes~\cite{lidmar03}, and
for a flat sheet in the linearized regime the parameters have the same
values as well.  If we
parameterized our model by spring constants $K\str$ and $K\bend$ equal
to the curvature of our radial and angular potentials at the bottoms
of their respective wells, we would have $\Ydisc = \frac{2}{\sqrt
3}K\str$ and $\kappadisc = \frac{\sqrt 3}{2}K\bend$.  In most cases
(except in direct comparisson with some experimental measurements), we 
are only concerned with the ratio of these elastic constants.  This ratio
provides a length scale, the \emph{Foppl-von K\'arm\'an length},
\begin{equation}
\label{eq:vkl}
       \vkl^2\equiv \kappabulk/\Ybulk.
\end{equation}
From this point on, we will take units such that $r_0=1$, and therefore
the parameter $\vkl$ is effectively dimensionless.

\later{Didn't we talk about incorporating a dimensionless
prefactor so that the exponential decay length of strain
near a disclination would  be $\vkl$?}

Previous work made this same ratio dimensionless using the capsid 
\emph{radius} $R$, rather than the triangle size $r_0$, and thus defined
the Foppl-von K\'arm\'an number~\cite{lidmar03}
\begin{equation}
  \vK = YR^2/\kappa.
\end{equation}
The capsid radius $R$ is well-defined in the case of a
spherical capsid, but for non-spherical capsids, a definition of $R$ is
problematic; and in any case, $\vkl$ controls many other properties,
such as the exponential decay of strain and Gaussian curvature with
distance from a defect.  Thus, we consider $\vkl$ to be the more
fundamental parameter and thus write $\vK = (R/\vkl)^2$.  We note that a
small $\vkl$ corresponds to a large Young's modulus and therefore
an \emph{angular} (or faceted) regime.  On the other hand, large
$\vkl$ entails a large bending stiffness and leads to a \emph{smooth}
regime \cite{lidmar03}.  Since our model is two-dimensional, we are
able to specify arbitrarily large $\vkl$.  Physically, however,
$\vkl$ must be on the order of the capsid thickness or smaller.
%\SH{You wanted me to cite something - I can't
%think of anything else, but I've already cited this one a bunch in the
%preceeding paragraphs}

\subsubsection{Spontaneous curvature and steric repulsion}

Microscopically, we expect that capsid proteins are more similar in
shape to cones or pyramids, with the apex toward the inside, than 
to cylinders~\footnote{Indeed, the CA protein in HIV consists of 
two separate domains, arranged in a wedge shape with the smaller 
C--terminal domain pointing towards the inside of the capsid and the
larger N--terminal domain on the outside.  See, for instance, the 
structures in \cite{vonschwedler03}.  RSV is similar~\cite{nandhagopal04}}.  
Therefore, if two proteins are in contact, the outer
surface will be bent at a characteristic angle.  This suggests that
$\ham\bend$ should favor some dihedral angle $\theta_0$, appearing in
Eq.  (\ref{eq:ham-bend}).  Additionally, it motivates our model of
steric repulsion based on tetrahedra, explained below.

The preferred dihedral angle $\theta_0$ is a key parameter since it is
the main determinant of capsid size in our model, as was speculated to
be the case in real capsids~\cite{caspar62}.  This corresponds to
spontaneous curvature in a continuum model.

The final term $\ham\steric$ in (\ref{eq:ham}) is a steric potential,
chosen to vanish for all physically realistic capsids.  The steric
potential proves difficult to incorporate into our cartoon model, for
two reasons.  Firstly, all the inter-unit interactions of properly
bonded units should already be accounted for in the elastic term
$\ham\str+\ham\bend$, so we demand that the steric force not make
additional contributions to these forces.  Secondly, the other terms
in Eq.~(\ref{eq:ham}) relate units that are ``topological neighbors'',
as defined by the bond network (the discrete configuration).  But two
parts of the capsid which are distant topologically may grow to be
nearby in real space (the positional configuration), and must then be
kept from intersecting.  Thus, the steric term must apply equally to
topologically distant segments of the capsid, or to adjacent units,
e.g. two as-yet unjoined triangles on the same vertex.

To implement a computationally tractable steric potential, we imagine
each triangle to be the base of an inward-pointing tetrahedron, and add
a repulsion between the apex of each tetrahedron and the vertices on
the base of each other.  This potential vanishes for physically
realistic capsids.  A more technical discussion may be found in
Appendix~\ref{app:steric}, and the steric Hamiltonian is defined in
Eq.~(\ref{eq:ham-steric}).

\subsubsection{Microscopic estimation of elastic energy}

Interactions between capsid proteins have been simulated
electrostatically~\cite{reddy98} to determine binding energies for
large multimers of capsid proteins, necessarily in different relative
positions.  Such simulations could be extended to determine the
elastic constants for particular viruses with known protein structure.

Alternatively, we can perform a rough estimate of the elastic
parameters by considering some experimental measurements.
\Citet{vliegenthart06} performed extensive computational studies to
relate experiments with atomic force microscope (AFM) indentation of
capsids to a model very similar to ours.  Thus, we can use these AFM
studies to determine the appropriate magnitude of $Y$ and $\kappa$.
\Citet{ivanovska04} carried out mechanical structure measurements on
the $T=3$ phage $\phi29$ and found the bulk modulus $B\approx
1.4\giga\pascal$ and the thickness $t\approx 2.5\nano\meter$.
% , and the radius $R\approx 25\nano\meter$.  
We obtain an estimate of the two-dimensional Young's modulus by 
$Y\sim Bt\approx 3.5\newton\per\meter$ \cite{nguyen05}.
% We can determine the size $r_0$ of each unit by comparing the surface
% area of a sphere to the area of $20T$ equilateral triangles, 
% $4\pi R^2 = 20T \frac{\sqrt{3}}{4}r_0^2$, so that 
% $r_0 \approx \frac{1.2R}{\sqrt{T}} \approx 17\nano\meter$.

We could also estimate the elastic parameters
from persistence length measurements.  \Citet{maeda85} measured the
tube-forming phage \fd{} in suspension and
determined the persistence length of the 
%$6.6\nano\meter$-diameter %(Fraden, many others)
$9\nano\meter$-diameter tubes at $22\degree\celsius$ to be $3.9\micro\meter$.  
If we construct a tube out of our triangular units, the persistence length 
would be
\begin{equation}
\label{eq:plength}
\xi_p \approx \frac{R}{k_BT}\left(\kappa+\frac{8}{9}YR^2\right),
\end{equation}
where $R$ is the radius of the tube.  Thus, we can conclude
$\kappa+(8/9)YR^2=Y\left((8/9)R^2+\vkl^2\right)\approx
22\electronvolt$, which puts an upper bound of
$0.17\newton\per\meter$ on $Y$, in sharp contrast to the $\phi29$
results above.  Moreover, since \fd{} is charged~\cite{zimmerman86},
the purely elastic contribution to the persistence length may be
much smaller, making our estimate very conservative.
If we previously knew either $\vkl$ or one of $Y$ or $\kappa$, we
could use this measurement to determine the others.

To get an idea of the elastic parameters for HIV, we can produce model
capsids by hand which resemble HIV cores.  In particular, we grew
several capsids with about 500 triangles in a cone shape.  Tuning the
elastic parameters to roughly match the observed shape of
HIV~\cite{benjamin05,vliegenthart06}, we found
$\gamma=(R/\vkl)^2\approx 550$ produces the correct amount of
facetedness.  This corresponds to $r_0/\vkl=6$.  Using our results for
capsid size as a function of $\vkl$ and $\theta_0$, presented in
Sec.~\ref{sec:size}, we can guess that such a capsid would require
$\theta_0\approx 20\degree$ to be grown by our model. \SH{Do you want
the comparisson with Nguyen to go in here as well?}

Given a set of connected triangles (a topological configuration), we can
now use this Hamiltonian to determine the lowest-energy configuration
of the positions of the triangles.  These positions correspond to a
continuous degree of freedom which is now fully determined by the
model ($\ham$) and the connectivities -- the discrete degree of
freedom.  Ultimately, we are only concerned with the discrete
configuration.

\subsection{Growth}
\label{sec:model-growth}

We have noted that capsids are determined by the locations of the
disclinations (pentamers).  For an irreversible growth model, in which
no step can be undone, the fundamental question is therefore: while
growth occurs at the border, which twelve vertices are frozen in as
pentamers?  Keeping this in mind, we will now discuss our capsid
growth process.

\subsubsection{Growth steps}

We define transitions between incomplete capsids, consistent with
irreversible growth, called \emph{growth steps}.  Two elementary growth steps
are immediately apparent: \emph{accretion} and \emph{joining}.  
Accretion is the
addition of a single triangle to a border edge and joining is the
formation of a bond between two adjacent border edges.  We require the
vertex between these two joined edges to have five or six triangles
around it in order to ensure that only pentamers and hexamers form.

\begin{figure}
  \includegraphics[width=0.45\columnwidth]{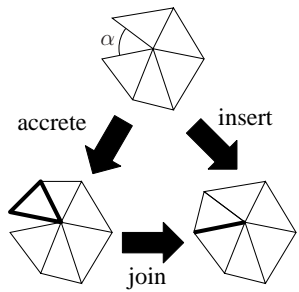}
  \caption{\label{fig:growsteps}Elementary growth steps of 
    \emph{insertion}, \emph{joining}, and 
      \emph{accretion}, performed from the same starting point
      of an edge with opening angle $\alpha$.
      Insertion can be decomposed into Accretion followed by Joining.}
\end{figure}

Besides accretion and joining, we define a third, composite growth
step: \emph{insertion}.  We define insertion as accretion followed by
joining along an edge of the new triangle.  The vertex into which we
insert must have four or five triangles.  Insertion at a 4-vertex or
joining at a 5-vertex is the only way to form a pentamer.  These three
steps are illustrated in FIG.~\ref{fig:growsteps}.

Growth begins with a small template -- either a single triangle or a
pentamer of five triangles about a vertex.  From here, the
growth is determined by the sequence of growth steps,
which is chosen stochastically.  We will first present our rules 
for the relative probability of choosing the growth steps, 
and then explain their microscopic rationalization.

\subsubsection{Rates}

\begin{figure}
  \includegraphics[width=\columnwidth]{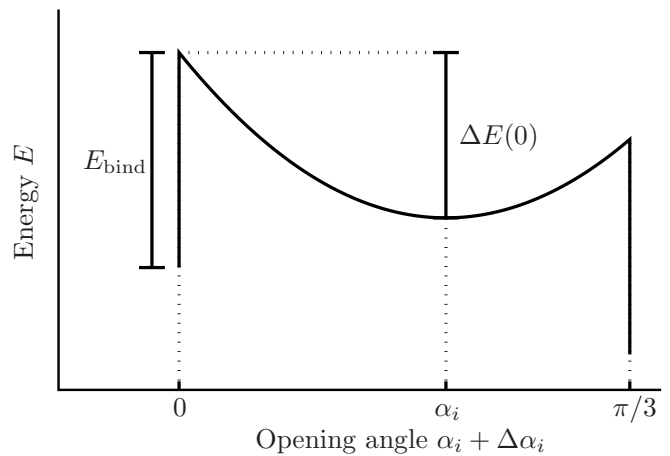}
  \caption{\label{fig:bindingenergy}A representation of the energy as a function
    of opening angle fluctuations $\Delta\alpha_i$.  When $\alpha_i+\Delta\alpha_i$
    reaches either 0 or $\pi/3$, we imagine joining or insertion, respectively,
    occuring, reducing the total energy by $E_\mathrm{bind}$, for joining,
    and some combination of $E_\mathrm{bind}$ and a chemical potential for
    insertion.}
\end{figure}

We precede each growth step by relaxing all vertex positions 
using a conjugate-gradient algorithm to minimize the positional energy
$\ham$.  Now a rate $k_\nu$ is defined for each allowed growth step
$\nu$, which is a function of the local topology and of the
\emph{opening angle} $\alpha$ between pairs of edges at each vertex on
the border, defined in FIG.~\ref{fig:growsteps}.  The probability of
step $\nu$ is then taken to be $k_\nu/\sum_\mu k_\mu$; once a step $\nu$ is
picked, we perform the step and iterate the process (beginning as
before with a relaxation).

We take the accretion rate, $k_A$, to be independent of the local 
configuration: in particular it is not a function of $\alpha$.

So long as we are concerned only with the outcome and
not the time taken to reach it, only relative rates are relevant. 
Thus we can now define
\begin{equation}
  \label{eq:kJ}
  \frac{k_J(\alpha)}{k_A} = \Gamma_J e^{-{\alpha}^2/2\sigma^2}
\end{equation}
\begin{equation}
  \label{eq:kI}
  \frac{k_I(\alpha)}{k_A} = \Gamma_I e^{-(\alpha-\pi/3)^2/2\sigma^2},
\end{equation}
with justification to follow.  Note that steps are only considered if
(1) they do not break any topological rules by enclosing a
non-pentamer/hexamer, and (2) they do not lead to steric hindrances.
This second point is discussed further in Appendix~\ref{app:steric}.

\subsubsection{Microscopic justification of rates}

While many models explicitly account for units in solution and
fluctuations in incomplete capsids~\cite{schwartz98,rapaport04}, we
have chosen a simplified cartoon model.  Implicit to this is the
idea that the capsid is thermally fluctuating between growth steps.

Say the time between successive additions is longer than the
relaxation time scale of the positional degrees of freedom.
Then between each growth step, we can assume that
the incomplete capsid is in equilibrium and thus samples a Boltzmann
distribution.  We consider the energy of fluctations about the relaxed
position.  For a particular vertex, $i$, with relaxed opening angle
$\alpha_i$, the energy of a fluctuation with opening angle
$\alpha_i+\Delta\alpha_i$ is well approximated by a quadratic, so that
\begin{equation}
  \Delta E(\Delta\alpha_i) \approx \frac{1}{2} A_i(\Delta\alpha_i)^2.
\end{equation}
We can therefore determine the elastic energy barrier for a vertex to
have an angle favorable for either insertion ($\alpha_i+\Delta\alpha_i
\approx \pi/3$) or joining ($\alpha_i+\Delta\alpha_i \approx 0$), and
thus the transition rates, $k_I(\alpha_i)$ and $k_J(\alpha_i)$,
respectively.  It is now clear that the rates defined above in
Eqs.~(\ref{eq:kJ})--(\ref{eq:kI}) are merely Arrhenius factors, with
\begin{equation}
\label{eq:sigma}
  \sigma^2 = \frac{k_BT}{A_i}.
\end{equation}
Note that $T$ here is temperature, and should not be confused with
the triangulation number defined earlier.

During any growth step, new bonds are formed.  We may consider an
extra energy term, $\ham\bind=-N_bE_b$, contributing a binding energy
$-E_b$ for each of the $N_b$ bound edges in the capsid.  Such an
energy is independent of the positional configuration.  For our
irreversible model to satisfy detailed balance, we need $E_b\gg\Delta
E(\Delta\alpha_i)$ so that the energy barrier for the reverse
transition is large compared to that for the forward transition.

The parameter $A_i$ and therefore $\sigma$ depends not only on the
elastic constants, but also on the local environment of the vertex in
the capsid.  We can determine normal values for $A_i$ by varying
angles on different capsids with different energy parameters, and
twice differentiating the Hamiltonian about the minimum.  Because most
of the opening angle fluctuations in physical situations are
\emph{in-plane}, $A_i$ depends most strongly on the Young's modulus,
and generally
\begin{equation}
\label{eq:d2H}
  A_i = \frac{\partial^2\ham}{\partial\alpha_i^2}\approx 0.1\Ydisc r_0^2.
\end{equation}
(Note that this is an absolute dependence on the energy scale
$\Ydisc$, and is the only reference we will make to an absolute
energy, since everything else depends only on the ratio
$\kappadisc/\Ydisc=\vkl^2$.)

We can perform a rough estimate of this width $\sigma$.  Using the elastic
parameters estimated for \fd{} in Sec.~\ref{sec:model-ham}, and assuming
$r_0\approx 4\nano\meter$, we find
$A_i \approx 0.1 \Ybulk r_0^2 \lesssim 17\electronvolt$.  We
therefore expect fluctuations of
\begin{equation}
  \sigma = \sqrt{\frac{k_BT}{A_i}} \gtrsim 0.038 \approx 2.2\degree
\end{equation}
%%%%%
at room temperature.  We will need $\sigma\gtrsim 10\degree$ for
satisfactory growth -- a reasonable possibility considering that we
conservatively ignored bending rigidity and charge.  Had we performed
this estimate using the much larger value of $Y$ from $\phi29$, we
would find fluctuations an order of magnitude smaller, leading to a
regime in which growth is not feasible.  But the $\phi29$ measurements
were taken from the head of a mature bacteriophage which is observed
to be much more faceted (small $\vkl$, large $\Ybulk/\kappabulk$) than
the immature form in which assembly occurs.  Such small fluctuations
are probably important for stability and infectivity, but also quite
detrimental to growth~\footnote{For a discussion of the geometric
considerations in phage head assembly, see
\bibinfo{author}{M.~F. Moody}, \bibinfo{journal}{J. Mol. Biol.}
\textbf{\bibinfo{volume}{293}}, \bibinfo{pages}{401} (\bibinfo{year}{1999})}.  
As such, we expect the immature
capsid to have much larger fluctuations, although no mechanical
studies have been done to allow this determination.

Sometimes a deterministic growth rule is preferred to the stochastic
rule presented above.  One possibility is a rule which accepts only
the move with the largest rate at any given point.

\section{Failure Modes}
\label{sec:fail}

The restriction that all vertices have either five or six triangles
can lead to problems in irreversible growth.  It is entirely possible
for a growing capsid in our model to perform a wrong growth step
resulting in a state which can never be completed -- that is, no
complete capsid satisfying the pentamer/hexamer-only requirement
includes the particular incomplete capsid in any of its possible
growth histories.  This section surveys two common failure modes. A
common theme is that the failure can be identified non-locally, long
before a step is reached at which the growth rules break down; a more
rigorous treatment is given in Appendix~\ref{app:comp}.

We cannot avoid considering failures, since we must exclude them
when reporting statistical distributions of the resulting
capsid ensemble (see Sec.~\ref{sec:results}).  More importantly,
we have taken for granted that actual physical assembly has a high
success rate (say, 10\% to 99\%).  Indeed, most of our labor 
on the project reported in this paper consisted of locating the
region of parameter space in which assembly had a high success rate.
Classifying the failure modes is a prerequisite to 
understanding what conditions reduce or eliminate then.

Failure modes are also experimentally pertinent.  Whatever the
``ideal'' capsid is for a given virus species, there is likely to be
more than one possible assembly model that produces it. But since
different models will tend to fail in different ways, they are better
distinguished experimentally by study of defective rather than of
ideal capsids.  If there are virus species which grow their capsids
near the limit of complete irreversibility, the resulting ensemble is
bound to contain mistakes.  Indeed, HIV cones have been observed
that are syrrounded with what is believed to be a second complete sheet
of capsid protein~\cite{benjamin05}.
%%%
\later{Find refs (Vogt?): ``Indeed, HIV is
known in some regimes to produce sheets or spirals and otherwise
proper capsid cones have been observed that are surrounded with what
is believed to be a second sheet of the protein-unit triangular
network~\cite{benjamin05}.''}

\subsection{Unfillable quadrilateral hole}
\label{sec:fail-quad} 

\begin{figure}
\includegraphics[width=\columnwidth]{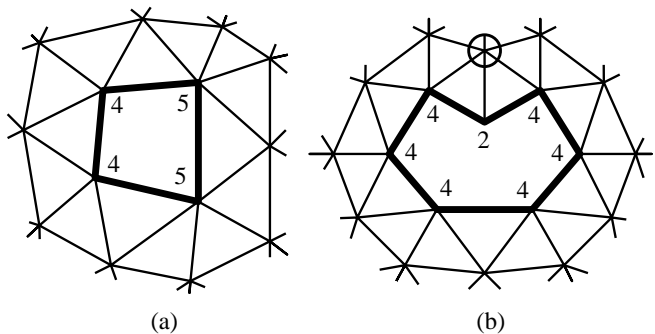}
\caption{\label{fig:hole}Incompletable holes at the end of growth.
Neither hole can be filled without introducing a heptamer.  The hole
in (b) would have been avoided had the circled vertex been made a
pentamer}
\end{figure}

First we look at a failure which occurs only at the end of a growth
process.  FIG.~\ref{fig:hole}(a) shows a common configuration with a
single quadrilateral hole.  Parallel transporting a vector around the
border gives no rotation and therefore there is no net disclination
inside (the net ``disclination charge'' is zero. -- recall the
discussion in Sec.~\ref{sec:model-conf}) The only conceivable filling
is with two triangles, but either possibility introduces a
7-coordinated vertex
\footnote{A 7-fold vertex has a negative disclination charge;
there must be a 5-fold vertex next to it, with its positive 
disclination charge so the interior of the loop is neutral. 
A pair or ``dipole'' of positive and negative disclinations
constitutes a \emph{dislocation}.  Its presence could have
been concluded by the non-zero Burgers vector associated
with that same loop; that is, the sum of the lattice displacements 
of each step, referenced to the ideal triangular lattice
that can exist far away from this hole}.

A less trivial example of this situation is shown in
FIG.~\ref{fig:hole}(b).  Here we can parallel transport a vector
around the border to see that a single disclination must reside within
the border; however, there is no way to fill in the remaining triangles to
satisfy this.  See Appendix~\ref{app:comp} for a more rigorous
discussion of this phenomenon.  If we continue growth, the hole will
eventually shrink to something similar to FIG.~\ref{fig:hole}(a).  
Some believe that such a hole is not detrimental to capsids, and in fact
capsids are suspected to be permeable to water and ions.
On the other hand, HIV is known to have a particle-to-infectivity 
ratio on the order of 100~\cite{piatak93}, and such holes, if they
are very common and detrimental to infectivity, may explain why
99\% of virions are not infectious.

This type of failure was common in all the growth rules we considered,
although it is more prevalent in certain situations.  In particular,
if the growth rate parameters defined in Eqs.
(\ref{eq:kJ})--(\ref{eq:kI}) are large, $\sigma\gtrsim 20\degree$ or
$\Gamma_{I,J}\gtrsim 200$, then creation of pentamers becomes very
random and is no longer based on the configuration.  In normal growth,
particularly at small $\vkl$ (angular regime), local strains cause
angles along the border to suggest whether a pentamer or hexamer
should be created, but large $\sigma$ decreases the sensitivity to
this.

\subsection{Crevice formation}
\label{sec:fail-crevice} 

\begin{figure}
\includegraphics[width=\columnwidth]{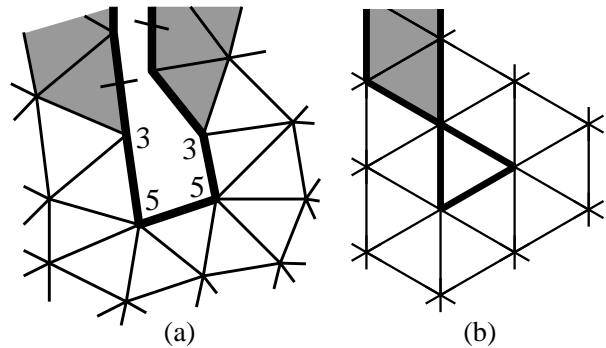}
\caption{\label{fig:crevice}(a) An incompletable configuration.  The
sequence of vertices with 3, 5, 5, and 3 triangles on a border can
never lead to a valid complete capsid.  Joining the marked edges would
produce an unfillable quadrilateral hole, similar to
FIG.~\ref{fig:hole}(a). (b) The same border flattened onto a
triangular reference lattice.  The shaded triangles from (a) now
overlap the corresponding triangles from the opposite side.}
\end{figure}

Next we look at a failure which can occur at any point during the
growth, called a \emph{crevice}. 
We see in FIG.~\ref{fig:crevice}(a) a portion
of a border with the four labeled vertices in a characteristically
incompletable configuration.  This becomes clear when the border is
flattened onto a reference lattice, as seen in
FIG.~\ref{fig:crevice}(b).  We now see that in the absence of
pentamers in the neighborhood of this section of border, several triangles
lie on top of others.  The introduction of a pentamer can only
make matters worse.  By effectively cutting out a 60\degree\ section
of the plane, it becomes even more crowded.  The only way to alleviate
this self-intersection is by introducing a negative disclination (heptamer).

Crevice failures can occur in different regimes, but arise in
particular during \emph{fingered} growth.  If accretions are much more
common than both joinings and insertions (such is the case when either
$\sigma\lesssim 5\degree$ or $\Gamma_{I,J}\lesssim 1$), then we expect many
long fingers only one or two triangles wide.  Crevices occur easily
between these fingers.  Even in the absence of fingers, sometimes
creating a pentamer will distort a neighboring vertex enough that the
angle is too large for insertion or joining.  This too often results
in a crevice.

Say a single crevice failure occurs during growth.  Further
growth outward from the failure should be prevented by steric
hindrance.  But growth elsewhere along the border will continue and
eventually fill in the crevice from its far end.  Then the capsid will
almost complete, leaving a small hole of the same type discussed in
Sec.~\ref{sec:fail-quad} above.  The two marked edges in
FIG.~\ref{fig:crevice}(a), for instance, might eventually join, leaving a
quadrilateral hole.

If two or more crevice failures occur, however, our model's resulting
capsid will have a network of cracks connecting these failures.
Real capsids might repair this problem by binding edges not sharing
a vertex (which is forbidden in our model); in
that case, the final result might instead have several small
quadrilaterals of the type seen in FIG.~\ref{fig:hole}(a).
% discussed in Subsec.~\ref{sec:fail-quad}.
%\CLH{later: tell the figure instead of saying that to identify?}

%A single crevice failure during growth will
%result in a small hole.  Two failures will result in a single linear
%crack in the final capsid.  If there are more than two failures, then
%the capsid will have a network of cracks.  This is an artifact of the
%steric repulsion and the lack of any binding other than directly
%connected neighbors.  Real capsids may indeed fix these problems by
%binding pairs which are topologically remote, although this leads to a
%risk of a mismatch if the bound units are different distances up the
%crevice.

%Another problem occurs when either $\sigma$ or $\Gamma_{I,J}$ is too
%large.  Since the first chance to form a pentamer is insertion at a
%four-triangle vertex, if this rate is significantly larger than the
%accretion rate even at unfavorable angles (80\degree\ or more), then
%we get uncontrolled pentamer formation.  At its extreme, this leads to
%$T=1$ capsid almost always.  There is a regime between this $T=1$
%growth and normal growth, however, in which the increased width
%prevents the detection of the correct position of the final one or two
%pentamers. \SH{I haven't yet talked much about how this detection happens}
%This is more of a problem at large $\gamma$

% look for 6-coordinated vertices with tiny R_join (a>45deg?) -> fail
% better incompletable test?
% sweep gamma and sigma at ~13-14deg at small/large gamma: classify T1,fail,normal

Although it is not as obvious, the smaller holes presented in 
Sec.~\ref{sec:fail-quad}
also have borders whose flattened images intersect themselves when cut at
certain places.  We can generalize this by stating that a border is
incompletable if there is any choice of cut which leads to
any triangles along the flattened border intersecting
one another.  The converse is true in most cases as well.

\subsection{Failure rates}
\label{sec:fail-rates}

If 5-fold vertices were simply incorporated at random moments during the
growth, virtually every capsid would fail in one of the two modes described
in  this section.  Since the topological constraints to be satisfied are
nonlocal, and the growth rates depend on local properties, it seems
mysterious at first how the growth  can be as successful as it is.
The key is that, in an elastic medium, the strain due to a defect
(such as a disclination) is also nonlocal;  at least, it decays
as a power law with the distance from the defect.  In this fashion,
the necessary information about the location of a faraway disclination
is passed to the growth border.

Since growth is stochastic, there is a possibility of errors
despite this passage of information.  All capsids are in danger of making
an error after the eleventh disclination is in place, and many are in
danger even earlier.

%%%%%%%%%%%%%%%%%%%%%

We can model the failure probability with a very simple assumption:
each time a triangle is added, there is a fixed probability $p_c\ll 1$
of starting a crevice.  This is not intended quite literally: $p_c$
must be understood as the fraction of edges along the border which can
possibly start a crevice, multiplied by the probability on each such edge that
this ``wrong'' step will be taken when a triangle is added there.
(The crucial step might be a ``joining'' but this contribution gets
folded in with the other one, since the border settles into a dynamic
near-steady state, so that the ratio of step types will be uniform on
averge.)

The survival probability of a defect-free capsid is thus 
\begin{equation}
   \frac{dP_\mathrm{sur}}{dN} = - p_c %dP_\mathrm{sur}
\end{equation}
where step $N$ plays the role of time, so that
\begin{equation}
   P_\mathrm{sur}(N) = P_0e^{-p_c N} .
\end{equation}

\CLH{I really ought to write $P_0 e^{-p_c N}$, since
there could be an initial transient while the cluster
is comparable to a lattice constant.  But I don't know
about the dependence of $P_0$ on $\theta_0$ or on $\vkl$.}

Growth will terminate after all twelve disclinations
have been incorporated, i.e. on average when $N=\Nbar(\theta _0,\vkl)$ 
(the mean size of capsids formed as a function of the parameters).
Furthermore, we hypthesize that $p_c \approx p_c(\vkl)$, i.e.
crevice formation depends strongly on the ratio of elastic constants 
almost not at all on the preferred angle $\theta_0$.
If so, the probability of success is 
\begin{equation}
  \label{eq:psucc}
  P_{\rm succ} = P_{\rm sur}(\Nbar) = e^{-p_c \Nbar}.
\end{equation}
Indeed, we will see the dependence of $p_c$ and $P_0$ on
$\vkl$ in Sec. \ref{sec:succ}

%%%%%%%%%%%%%%

\SH{We can do similar analyses on pentamer rates, depending on $\theta_0$
mostly in the beginning, history in the middle, and then $\vkl$ at the end}

\section{Results}
\label{sec:results}
Here we discuss several measurements which can be used 
to quantitively characterize various properties of 
capsids (individually, or as an ensemble) specified by
a triangulation of vertices, such as the results of
our growth model.  Our results fall into three general
categories: size, success, and shape.  First we look at
the size of the resulting capsids and show the dependence
on the elastic parameters.  Next we look at the probability
of successful growth, in terms of both the size of the
capsid and of the growth rate parameters.  Last we
comment on measures of capsid shape which, along with
capsid size, is a measurement which can be used with
data from cryo-EM experiments.

FIG.~\ref{fig:capsid} shows an example of a typical capsid shell
resulting from our growth simulation.  This capsid emphasizes that our
configurations are inherently random and irregular.  The degree of
``lumpiness'' in the external shape depends strongly on the Foppl-van
K\'arm\'an length $\vkl$, as is elaborated in Sec.~\ref{sec:shape},
below.

Each capsid is grown until either a successful completion, or an
identifiable failure, such as a self-intersection in the flattened
border.  Relaxations are minimized until the gradient-squared is less
than $10^{-6}$, in units with $K\str\gtrsim K\bend=1$.  The entire
growth process for a small capsid takes several minutes on a 1.6GHz
processor, while a large capsid takes many hours, the majority of the
time devoted to minimizing energy.  The following plots of size and
success rate include data from \SH{change this if new data comes in}
134,352 capsids.

%\todo{Here or in rhis section, please give some statistics as to how
%many runs were done, how long it takes on a xxx GHz processor,
%how many iterations were used (and what stopping criterion)
%for the relaxation, etc.  Does most of the run time go into 
%relaxation steps?}

\subsection{Size}
\label{sec:size}
%\CLH{If you want to say energy, say ``free energy'', but $\theta_0$
%isn't an energy.}

The simplest thing to observe about a capsid is its size.  We can
count the number of triangles $N$, or measure the average radius $R$.
%Even for incomplete capsids, a least-squares fit to the surface of a
%sphere can yield a radius of best fit. 
As expected \cite{caspar62,bruinsma03}, capsid size depends most 
heavily on two parameters from our effective 
Hamiltonian, $\vkl=\sqrt{\kappabulk/\Ybulk}$ and $\theta_0$,
which we rewrite as a length
\begin{equation}
  \label{eq:scl}
  \scl \equiv \frac{r_0}{2\sqrt{3}}\cot(\theta_0/2).
\end{equation}
This length is the radius of curvature from two equilateral triangles
with side length $r_0$ joined at an angle $\theta_0$ and tangent to a
common sphere.  We now have three length scales, $r_0$, $\scl$, and
$\vkl$.  It is useful to think of these as two dimensionless parameters,
$1/\scl$ and $1/\vkl$, taking $r_0=1$.

In the smooth regime, when $1/\vkl\lesssim 50/\scl$, the
variation in the dihedral angles at different bonds is small, so the
radius of the resulting capsids tends to $\scl$.  For larger $1/\vkl$
(the angled regime), the Young's modulus increases.  Hexamers, which
make up most of the capsid, become flatter.  Thus, the effective
preferred angle $\theta_0^\mathrm{eff}$ decreases, resulting in larger
capsids.
\SH{Additionally, the self-energy of each disclination increases,
which may also increase capsid size~\cite{bruinsma03}.}
\SH{CLH comment to discuss, but this isn't a strong statement, since
it depends on (a) \theta_0=0, and (b) a ``switching'' picture, to make
sense.}

\begin{figure}
  \includegraphics[width=\columnwidth]{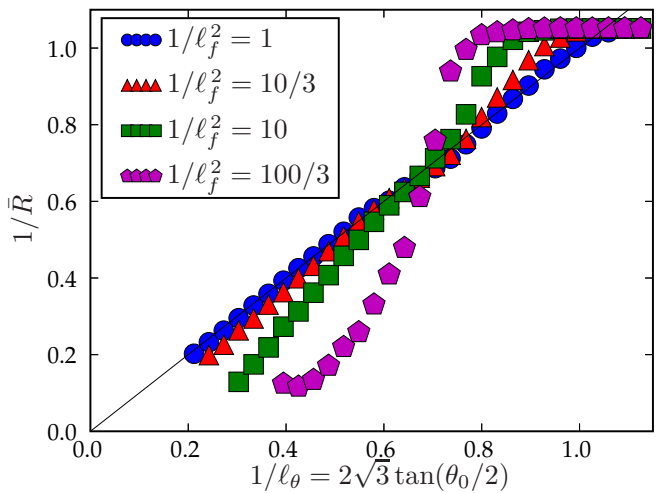}
  \caption{\label{fig:size}(Color online) Plot of $1/\Rbar$ versus 
    $1/\scl = 2\sqrt{3}\tan\theta_0/2$, in units of $r_0$.  We see that 
    in the smooth regime of small $1/\vkl$, the mean capsid radius $\Rbar$
    very nearly follows $\scl$.  In the angled regime (large $1/\vkl$), 
    we find smaller capsids (many $T=1$) for a much larger range of
    $1/\scl$, followed by a sharper increase in size at smaller cunrvatures.
%    \SH{cut out -- The $1/\vkl^2=100/3$ curve suggests that this linear regime ends at some
%      point, and should begin to approach the origin.}
    Parameters with fewer than 10 successful capsids were omitted.}
\end{figure}

\begin{figure}
  \includegraphics[width=\columnwidth]{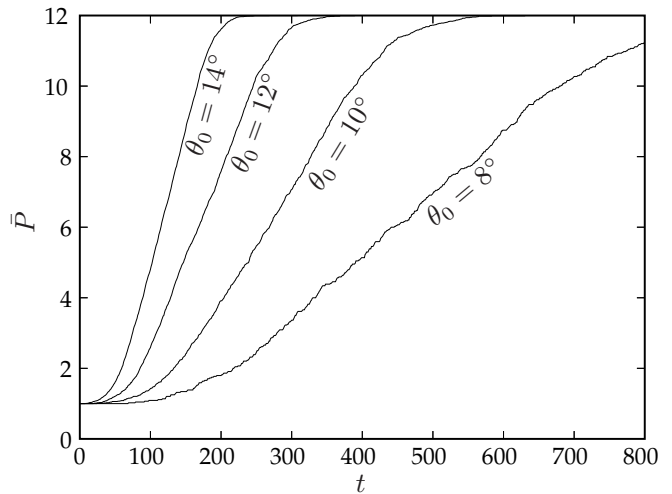}
  \caption{\label{fig:pent}Plot of the average number of pentamers
    $\Pbar$ versus the step number $t$, which is nearly equivalent to
    the number of triangles $N$.  This gives a picture of the general
    pathway of growth behind the curves in FIG.~\ref{fig:size}.  This
    growth was carried out at $1/\vkl^2=10/3$, and different
    spontaneous curvatures $\theta_0$ as shown in the figure.  We see
    that growth consists of an initially slow process to add the
    second pentamer, followed by a rather linear regime in which
    $d\Pbar/dN$ is roughly constant.  Note that both the initial rate
    at $P=1$ and the following slope depend on $\vkl$.}
\end{figure}

We simulated many capsids assembling at four values of $1/\vkl$ and
$\theta_0$ between 7\degree\ and 36\degree.  For each set of parameter
values, we averaged the radius of the completed capsids, $\Rbar$, and
plot the inverse of the radius $1/\Rbar$ as a function of the
spontaneous curvature $1/\scl = 2\sqrt{3}\tan(\theta_0/2)$.  in
FIG.~\ref{fig:size} for several different values of $\vkl$.  We see
that, for large $\vkl$, the curves roughly follow the line 
$1/\Rbar=1/\scl$.  As $\vkl$ decreases, we see a very different
behavior, which favors small (mostly $T=1$) capsids for a much
larger range of $\theta_0$, before the size suddenly increases
very quickly around $1/\scl\approx 0.7$.
%distance from this line increases as $1/\vkl$
%increases.  Notice that we can clearly see two different regimes in
%this plot.  The $(1/\vkl)^2=10$ curve in particular seems to change
%behavior near $\Rbar\approx 4r_0$, corresponding to
%$\gamma=(R/\vkl)^2\approx 160$.  This is suggestive of a buckling
%transition \cite{lidmar03}.
We can see what is behind these curves in FIG.~\ref{fig:pent}, which
shows the average growth history for several individual parameters,
represented by the average number of pentamers $\Pbar$ as a function 
of the number of growth steps $t$.

%\SH{remove this whole paragraph:}
%\SH{Because most of the larger capsids failed, we used the best-fit radius
%for any capsids with at least 9 pentamers at the time of failure.  Also,
%very small ($<100$ triangles) capsids were removed from the averages for
%$(r_0/\vkl)^2\ge 100$, because they were an artifact of too large
%a width $\sigma$ in the growth rules. \later{add hook - as discussed
%in next section}}

\begin{figure}
  \includegraphics[width=\columnwidth]{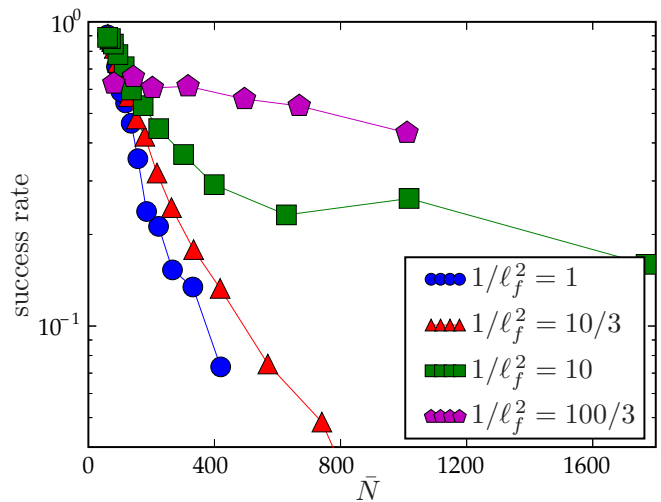}
  \caption{\label{fig:succ}(Color online) Plot of success rates as 
    a function of $\Nbar$ for the given parameters, from the size
    measurements.  We see a somewhat exponential decay, suggesting 
    that introduction of errors is a Poissonian effect, as discussed in 
    Sec.~\ref{sec:fail-rates}.}
\end{figure}

\begin{figure*}
  \includegraphics[width=2\columnwidth]{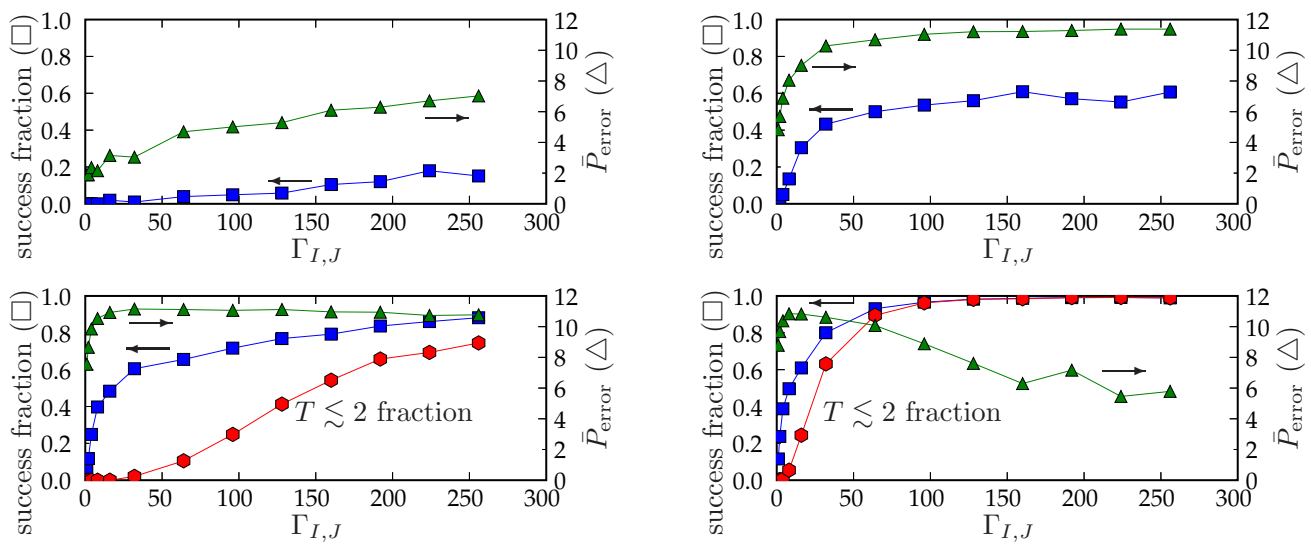}
  \caption{\label{fig:rsucc}(Color online) Success fractions as a
    function of rate parameters $\Gamma_{I,J}$ and $\sigma$
    ($\sigma=8\degree$, top left; $\sigma=12\degree$, top right;
    $\sigma=16\degree$, bottom left; $\sigma=20\degree$, bottom right)
    The square markers show the fraction of successful capsids at each
    parameter, including small capsids.  For $\sigma=16\degree,
    20\degree$, the hexagons mark the fraction of total capsids which
    were small (successful and less than 46 triangles, or
    pseudo-$T\lesssim 2$).  The smaller values of $\sigma$ had no such
    capsids.  Finally, the triangular markers show the average number
    of pentons $\bar P_\mathrm{error}$ at the time of error detection.
    Note that $\bar P_\mathrm{error}\to 12$ means that all the errors
    are small holes at the end of growth.  \SH{note sigmoid-like shape
    of $T=1$ curve}}
\end{figure*}

\subsection{Success rate}
\label{sec:succ}
An important consideration for an irreversible growth model is under
what circumstances it successfully produces complete capsids.  We have
already shown that a variety of failure modes exist, resulting in
incompletable capsids.  We can easily quantify how often these
failures actually occur as a function of parameters.  We predicted in
Sec. \ref{sec:fail-rates} that the success rate should be exponential
with the expected size of the capsid.  For each choice of parameters,
we average the sizes (measured by number of triangles, in contrast
to radius as in FIG. \ref{fig:size}) of the capsids, and thus
map the parameters $\theta_0$ and $\vkl$ to $\Nbar(\theta_0,\vkl)$.
We then plot the percentage of capsids that completed successfully when
grown with these parameters in FIG. \ref{fig:succ}.  While there is
systematic deviation from exponential decay, due to the many 
considerations left out of our analysis, we do still see a mostly
exponential trend in the data.

We see in FIG. \ref{fig:succ} that for a given size, growth is
generally more successful for more faceted capsids.  For large capsids
(best-fit radius $\bar r\gtrsim 10 r_0$), the failures in the smooth
regime ($1/\vkl^2\le10/3$) all occurred in the early stages of growth,
in which only a few pentamers had been added.  This suggests that
large bending stiffnesses lead to more common crevice failures.  On
the other hand, the faceted capsids ($1/\vkl\ge10$) failed mostly in
the late stages, in which only several pentamers were missing,
suggesting that faceted capsids are somehow resistant to crevice
failures and instead fail with small holes.

In Sec.~\ref{sec:fail}, we mentioned the impact of the rate
parameters $\Gamma_{I,J}$ and $\sigma$ on successful completion.  
We measured the failure rate as a function of these parameters, 
using reasonable values of $\theta_0=16\degree$ and $\vkl^2=0.1$ In
FIG.~\ref{fig:rsucc} we plot the fraction of failed capsids
due to either small holes at the end of growth, or crevice failures in
the middle of growth.  We see that small values of
$\Gamma_{I,J}$ and $\sigma$ indeed produce errors.  Larger values of
$\sigma$ and $\Gamma_{I,J}$ produced successful capsids, but almost
all were $T=1$.  This particular result is very sensitive to our
particular growth rules, and a choice which prevented insertion until
there were five triangles around a vertex would drastically change
the result.

%\begin{figure}
%\SH{Figure showing success vs. $\Gamma$ and $\sigma$}
%\caption{\label{fig:success_rates}Fraction of successful capsids as
%a function of rate parameters $\Gamma_{I,J}$ and $\sigma$.}
%\end{figure}

\subsection{Shape}
\label{sec:shape}
Beyond size and success, most other measurements fall under the
category of shape measurements.  In particular, we might measure either
the degree of symmetry or the facetedness of a capsid.

Spherical harmonics may be useful for evaluating icosahedral symmetry,
as spherical harmonic coefficients of icosahedrally symmetric functions
vanish for all but $\ell=0,6,10,\ldots$.

\Citet{kingston01} uses the asphericity, defined as the ratio of
inradius to circumradius to measure the faceted shape of RSV capsids.
\Citet{lidmar03} also defined an asphericity, $\langle R^2\rangle/\langle
R\rangle^2$.  While these are good measurements for symmetric capsids,
they are not useful for the irregular capsids we grow, because
they cannot distinguish between, for instance, a smooth egg-shaped capsid
and a faceted spherical capsid.  We instead use a measure based on the
Gaussian curvature, described below.

\subsubsection{Curvature}
\begin{figure}
  \includegraphics[width=\columnwidth]{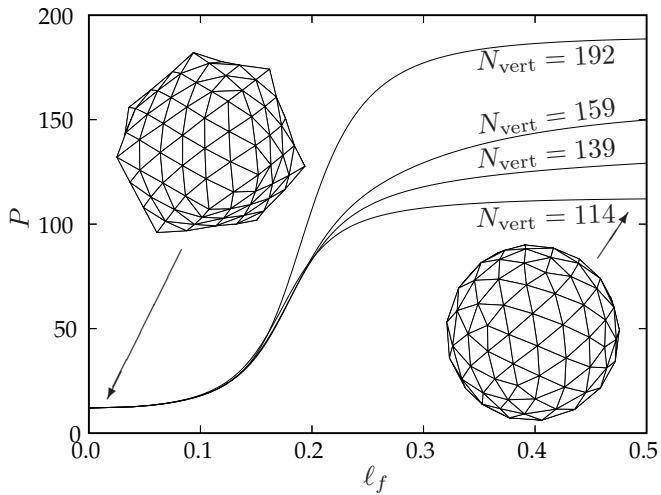}
  \caption{\label{fig:ipr}Inverse participation ratio for four capsids
    with different numbers of vertices $N_\mathrm{vert}$.  We can see
    that at $\vkl=0$ (the angular limit), $P=12$, and as
    $\vkl$ increases (the smooth limit), $P\to N_\mathrm{vert}$.  The
    exact shape of the curve depends on the placement of pentamers,
    but in general we see an inflection point around $\vkl=0.18$,
    which corresponds to $\gamma=R^2/\vkl^2$ roughly between 200 and
    400.}
\end{figure}

In light of recent advances in tomography, a very relevant measure is
Gaussian curvature $K$.  In our discrete triangular model, we can
measure the integrated Gaussian curvature $I=\int K\,da$ over the
neighborhood nearest to a single vertex by measuring the area
(equivalently, angle surplus) of the spherical polygon traced out by
the incident triangles' unit normals.  We can easily extend this to
the integrated curvature over all the vertices within any loop around
the capsid.  The integrated curvature over the entire capsid is always
$4\pi$, a topological invariant related to the Euler characteristic.
The question then arises how this curvature is distributed over the
capsid.  For highly faceted capsids, each pentamer has $I\sim\pi/3$,
while the rest of the capsid has $I\to0$.  On the other hand, the
curvature is distributed uniformly over smooth capsids.  This
motivates the definition of an inverse participation ratio (IPR),
\begin{equation}
  \label{eq:ipr}
  P = \frac{(\sum_j I_j)^2}{\sum_j I_j^2} = \frac{(4\pi)^2}{\sum_j I_j^2},
\end{equation}
where $I_j$ is the integrated curvature about vertex $j$.  This
essentially measures the number of lattice sites the curvature is
localized to.  The IPR is plotted for a single capsid relaxed to
different elastic parameters in FIG.~\ref{fig:ipr}.  We see that
$P=12$ at $\vkl\to 0$ while $P\to N_\mathrm{vert}$ at
$\vkl\to\infty$.

\begin{figure}
  \includegraphics[width=\columnwidth]{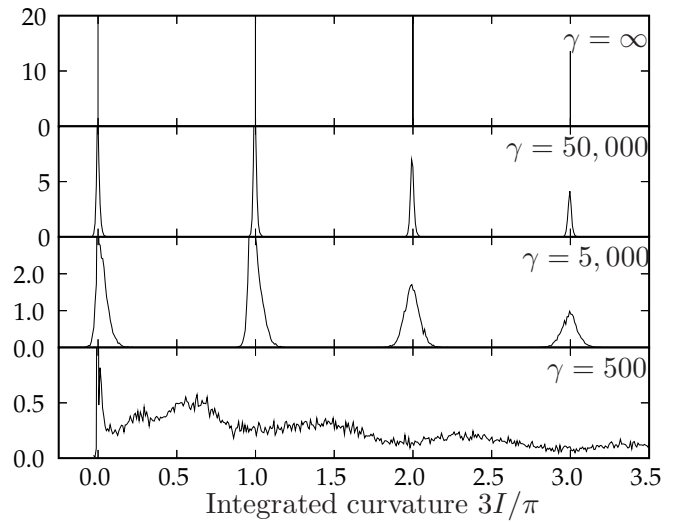}
  \caption{\label{fig:curvdist}Distribution of integral curvatures
    within random loops around random capsids relaxed to four different
    elastic parameters, characterized by $\gamma=(R/\vkl)^2$.  The 
    distribution is sharply peaked at the integers for the angled 
    regime at small $\vkl$ and diffuse for the smooth regime at large 
    $\vkl$.}
\end{figure}

This same integrated curvature can be measured on triangulated
tomographical data from capsids.  The integrated curvature within
large loops should be relatively stable even if the Gaussian curvature
varies quickly.  For an arbitrary loop around a capsid, we will get a
contribution of $\pi/3$ from each enclosed pentamer.  The loop may
then be pulled tighter to pinpoint the location of each pentamer.  We
simulated this process by growing a large number of random capsids and
integrating the curvature within many random loops on each.  Each
capsid was relaxed to several different values of $\vkl$.  The
resulting distribution of curvatures is displayed in
FIG.~\ref{fig:curvdist}.  At large $1/\vkl\approx 20$ we see very sharp
peaks.  These peaks diffuse into a mostly uniform background by
$1/\vkl\approx 2$.

\subsubsection{Average dihedral angle}
\begin{figure}
  \includegraphics[width=\columnwidth]{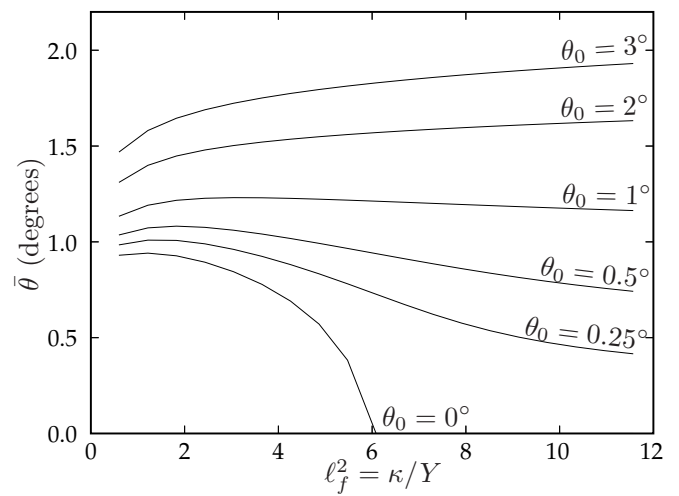}
  \caption{\label{fig:bendless}Average dihedral angle of a large
    pentagonal sheet with a single disclination in the center.
    Plotted versus $\vkl$ at different $\theta_0$.  The bottom
    $\theta_0=0$ curve shows a first-order buckling phase transition.
    Each subsequent curve increments $\theta_0$ by 0.25\degree.}
\end{figure}

We can measure the average dihedral angle of either a growing
or a complete capsid.  FIG.~\ref{fig:bendless} shows a graph
of the average dihedral angle for a very large pentagonal sheet
with a single disclination in center as a function of $\vkl$,
aat different $\theta_0$.  We see a first order phase transition at
$\theta_0=0$.
%and curves generally reminiscent of $\langle M\rangle$
%versus $T$ plots of the Ising model at different $H$.

%\subsection{Other measurements}

%\subsection{Growth history}

%\subsection{Symmetry}

\section{Discussion}

In this section, we recapitulate the highlights of our model and
the simulation results, and outline extensions that could improve 
their realism.

\subsection{Summary of results}
\SH{You say this seems to jump topics?  I don't see it.}
Our irreversible growth model, based on trimer units with the simplest
possible Hamiltonian and growth rates, did succeed at producing closed
capsids, but only when the parameters are tuned to the proper range:
$1/\vkl\lesssim 10$, $\theta_0\gtrsim 8\degree$, $\Gamma_{I,J}\gtrsim 50$,
and most importantly, $\sigma\approx 12\degree$.
%\CLH{Is it really a fine tuning?  My impression was you had freedom of
%an order of magnitude.  That's different from a parameter that can
%only vary by (say) 20\%, which I thought was the case for Nguyen's
%result.}

Our model (Sec.~\ref{sec:model}) made a sharp division between
configuration variables that were continuous (position) and discrete
(bonding topology); correspondingly the model parameters were divided
between a Hamiltonian (harmonic form) and rate constants for a set of
first-order processes; for simplicity, monomers in solution were not
treated explicitly.  The model's most distinctive feature is its use of
trimer units (triangles), which turns out to have several inherent
disadvantages.  First, our ``insertion'' step
(Sec.~\ref{sec:model-growth}) seems as though it should be been
redundant; unfortunately, omitting this (relying on the ``joining''
step in its stead) produces abundant failures -- the fingered growth
and crevices elaborated in Sec.~\ref{sec:fail-crevice}.  In other
words, good growth depended on joinings being rare compared to
insertions, which followed from our growth rules
(Sec.~\ref{sec:model-growth}), since opening angles $\alpha$ near
0\degree\ are much less common than those near 60\degree.  This may
mean that if a capsid assembles from trimers in solution, the
only way to have normal growth is that there must be cooperative
binding as in our insertion step.

Our results may be divided into two categories: the growth process
(including the success rate) and the shape of the resulting model
capsids.  In the first category, we found mathematical descriptions
(Appendix~\ref{app:comp}) which clarified the constraints on the
positions of the fivefold-coordinated vertices which fully
characterize the bond network.  Additionally, we uncovered a simple
relation between the chance of failure and capsid size,
Eq.~(\ref{eq:psucc}).  In the second category, we showed the
relationship between the capsid's final size and the two length
parameters, $\vkl$ and $\scl$.  In contrast to \citet{nguyen05}, we 
note that the \emph{average} capsid size is indeed well-determined 
by the spontaneous curvature parameters for large capsids, even in the
absence of any scaffolding considerations.  We also
extended the concepts of Ref.~\cite{lidmar03} to irregular capsids.
In particular the ratio of bending and stretching stiffnesses -- which
we suggest is best parametrized by a length, Eq.~(\ref{eq:vkl}),
rather than a dimensionless ratio -- controls whether the resulting
shape is smooth or angular, as we have characterized by an inverse
participation ratio, Eq.~(\ref{eq:ipr}).

One application of our results from different $\vkl$ relates to
phage maturation.  Many phages include a maturation step in which
the assembled prohead greatly increases its stretching stiffness
relative to its bending stiffness, making it more faceted.  One might
ask why the assembly process occurs in the smooth regime, especially
since our results in Sec.~\ref{sec:succ} show that the probability
of success is smaller in this regime.  We propose that the advantage
of growth in the smooth regime is size selection.  In FIG.~\ref{fig:size}
we see that faceted capsids have a sharp transition in size around
the region most phages fall into ($T=3$ to $T=7$).  As might be expected,
for a given set of parameters, the spread of capsid sizes is also much
broader near this transition.  Thus, in order to well-control the
size of assembled capsids, a virus might prefer to grow in the
smooth regime, counting on other factors such as scaffolding to increase
its chances of successful assembly.

\subsection{Future directions: more realistic random growth}

\subsubsection{Models with non-trimer units?}

The retroviral CA proteins we claim to model have well-documented
dimerization~\cite{delalamo05a} and hexamerization~\cite{mortuza04}
interactions, but no trimer interactions have been
observed in retroviral capsids.  A model based instead on pentamers
and hexamers could be implemented simply by changing the growth steps
to add several triangles at a time, so as to fully enclose a single
vertex each step into either a hexamer or a pentamer.  
%\CLH{Too vague?  ``A more thorough implementation would use the
%honeycomb lattice (dual to the triangular) , defining energies more
%appropriate to pentagons and hexagons of this lattice.''  Just which
%terms of Ham.?} \SH{All I can figure is polygon area, switching cost,
%and I've tried to think of some way to get the Gaussian curvature,
%since these hexagons and pentagons are dual to our vertices, and the
%vertices are really the ``carriers'' of curvature in our model -- so I
%have a feeling that the polygons of the dual model should have a
%similar property, but I don't see a good way to do it.}
We gain some benefit, however, from actually changing our
representation to a honeycomb lattice -- the dual to our current
triangular lattice.  Vertices of the dual lattice are all
three-coordinated, so each vertex along the border has either one or
two capsomers attached to it -- much simpler than the five different
possible coordination states for border vertices in the trimer model.
In this model, growth rules could explicitly depend on the total
coordination of a vertex.  Such coordination-based rules greatly
assisted successful growth in our trimer model, but were not as
physically justifiable as they are in the dual model.

These considerations suggest that behavior arising from this choice
\SH{``this choice'' is circled -- why?}
is not universal.  We expect models based on dimers, trimers, or
pentamers and hexamers to fall into different universality classes.

Another direction leading to a more realistic model is to improve the
accuracy of our interactions.  Microscopic electrostatic simulations,
such as with the CHARMM software \SH{CLH comment about needing to say
what the software is for -- I think I do that sufficiently already}, 
could provide a more realistic
Hamiltonian for specific viruses, which could be included in future
models.

\subsubsection{Lattice fluctuations}

%\CLH{Maybe clip this, at least till further discussion.  Not devloped
%enough here.}  \later{Another model worth consideration is one based
%on Langevin dynamics.  Rather than assume the incomplete capsid fully
%relaxes between each growth step, we can simulate brownian motion and
%define criteria for accretion and binding based on local geometry and
%time constraints.}

A deeper understanding of the relationship between topological
configurations is critical.  So far we have only thoroughly considered
irreversible growth transitions.  Other transitions relate to the
motion of disclinations on the lattice (always in pairs), both for the
purpose of ennumerating the near-symmetric states, and for an
understanding of the rearrangement dynamics by which real capsids may
anneal their bond configurations into the free energy minima predicted by
many equilibrium models.

\SH{too terse?}

\subsection{Future directions: realistic shapes}

All well-studied real capsids exhibit greater regularity than
our current model can regularly generate.  
How can the Hamiltonian (or the growth dynamics) be
modified so as to generate an icosahedral, or (for HIV)
conical capsid?

\subsubsection{Icosahedral symmetry}

The main challenge for theory is to explain 
the assembly of icosahedrally symmetric capsids, 
if one is not close to equilibrium.
Hamiltonians such as ours do indeed give effective
repulsion between the disclinations, and the 
free energy minimum is known to have icosahedral symmetry
in similar models~\cite{bruinsma03,lidmar03}.
However, this is simply insufficient to produce large
symmetric capsids in a model where the accretion rate
depends on local geometry, 
since the growing border does not contain enough information in 
just the opening angles (Sec.~\ref{sec:model-growth}).
Even deterministic variants of the growth model
never yielded icosahedral capsids larger than $T=4$.

We speculate that if the bending potential $\ham\bend$ was not simply
harmonic around $\theta_0$, but instead had minima at two different
angles $\theta_1$ and $\theta_2$, this might robustly favor a regular
pattern of edges with $\theta_1$ and $\theta_2$, thus permitting
determination of larger icosahedral capsids.  A double-well potential
would presumably represent some sort of conformational switch, perhaps
an internal bending between two domains of the capsid protein.  Thus,
this proposal has some features in common with the matching-rule
models that we dismissed as implausible (Sec.~\ref{sec:intro-models}),
but anharmonic potentials seem much more natural than variations in
the edge-binding (which, in our model, corresponds to the term
$\ham\bind$ mentioned briefly in Sec.~\ref{sec:model-growth}).

One other change which could result in more symmetric capsids, as well
as more successful growth in general, is to relax our irreversibility
constraint.  Allowing the growing edge to ``melt back'' would allow a
growing capsid to better explore the possible configurations, in
particular curing crevice and fingering defects.

\subsubsection{Retroviruses}

We asserted that the randomness of our model's growth behavior makes
it appropriate for modeling the irregularity and pleiomorphism
observed in the capsids of retroviruses such as HIV.  However, mature
HIV capsids do have a typical gross shape, which is mostly conical
(although sometimes tubular) in vivo, whereas our current model grows
round capsids on average.  A cone is charaterized by having (say) five
disclinations around its smaller end, seven around the large end, and
none on the belt between; this means the rates of adding pentons must
somehow vary during different stages of the growth.  When cones form
inside an envelope, the difference could be attributed to depletion of
the monomers as they are incorporated into the capsid: that (see
Sec.~\ref{sec:model-growth}) would decrease the rate of insertion but
not of joining, leading to a greater chance of penton formation.  A
difficulty with concentration control is that cone completion leaves
in solution 70\% \cite{briggs04} of the capsid proteins: in order for
this to grossly affect the rates, accretion must microscopically be a
rather high-order process. It also leaves unexplained the large
density of pentons at the {\it earliest} stage: a possibility is to
add a simple interaction between the capsid and either the nucleic
acid or the membrane \cite{benjamin05,briggs06}.

\begin{acknowledgments}
We would like to thank John Briggs for access to his unpublished work,
and Robijn Bruinsma, Grant Jensen, Diana Murray, and particularly Volker 
Vogt for many discussions.  Our interest originated in the suggestion of
B. Shraiman, who pursued similar growth simulations on the aggregation
of clathrin cages~\cite{shraiman97} and on virus capsids~\cite{socci00}.
This work was supported by U.S Dept of Energy grant No. DE-FG02-89ER-45405.
\end{acknowledgments}

\appendix
\section{Completability}
\label{app:comp}
It is possible to grow an incomplete capsid which is not part of any
allowed capsid.  
This appears to be a consequence of our rule that
a capsid vertex can only have coordination 5 or 6.  
(Seven-coordinated vertices, were they allowed, would
let the capsid recover from almost every 
``mistake'' discussed in this section.)

As a complement to the more qualitative discussion in
Sec.~\ref{sec:fail},
this appendix presents the technical criteria we 
discovered to identify when a partial capsid is or is not
completable, non-locally and long before the growth 
rules carry us to a point where we must make a 7-fold
vertex or stop.
The completability conditions are defined entirely in terms of the 
%%% topology of the 
growing  border, which can be uniquely described
by traversing the vertices (in a specified direction) and
listing the number of triangles present at each vertex.  Thus
a string of numbers from 1 to 5 specifies a border.  (6 is allowed,
but is trivial.)

\subsection{String representation}
We can represent any border by a word $a_1a_2\ldots a_n$, where $1\le
a_i \le 6$ is the number of triangles around the $i^{th}$ vertex,
counting clockwise from an arbitrary starting point.  We define
several operations on these string representations.
First, consider
\begin{equation}
A(a_1a_2\ldots a_n) \equiv 1(a_1+1)a_2\ldots(a_n+1)
\end{equation}
and
\begin{equation}
J(a_1a_2a_3a_4\ldots) \equiv (a_1+a_3)a_4\ldots,
\end{equation}
representing accretion and joining, respectively.  Note that the $a_2$
term disappears upon applying $J$.  This vertex is enclosed and is no
longer part of the border.  We therefore require $a_2=5$ or 6.  We can
further define insertion $I = J \circ A$ as the composition of joining
and accretion.  Finally, because we defined these operations to act on
the starting and ending points of our string representation, we must
define a cycle operation, $C(a_1a_2\ldots a_n) \equiv a_2\ldots
a_na_1$.  Because of the unimportance of the starting point in
representing a border, cycling leaves borders invariant.  Since
these operations are sufficient to grow any capsid, we can uniquely
describe a capsid by the sequence of operations on the border required
to arrive at the border from a single triangle, $111$.

Using this representation we can immediately identify some borders
which are incompletable.  Consider the border $X=555\ldots$
Joining is illegal since it leaves a
vertex with 10 triangles.  Accretion leads to $A(X)=6166\ldots$
which clearly cannot be completed since only joining can be done
on the 6's, and this leaves seven triangles about at least one
vertex.  Finally, insertion yields $I(X)=66\ldots$ which is
incompletable for the same reason.

Any border which intersects itself on a flat reference lattice is
incompletable (coincident edges are allowed).  It is important to take
notice of which side of the border is the inside (from which the
triangles are being counted) and which is the outside.

We thus define the complement of a border 
\begin{equation}
\overline{a_1\ldots a_n}\equiv (6-a_n)\ldots(6-a_1).
\end{equation}
If the original border enclosed $d$
disclinations then its complement encloses $12-d$ and can be glued
together to form a complete capsid.  We must note two things.  Firstly,
the complement of a border may be a border which cannot possibly be 
grown using our growth operations.  Secondly, the complement is only
unique insofar as the seam between the two incomplete capsids is occupied
only by six-fold vertices.  However, many ``pseudo-complements'' may
be constructed which leave disclinations on this seam.

While the border by itself is useful for analyzing completability, it
does not uniquely describe the interior.  An individual border may 
have many different realizations, with disclinations in different 
positions.  In fact, a pair of disclinations can move in opposite
directions (relative to a common reference lattice, if one exists) 
without changing the border.

\subsection{Winding number}
We can compute the winding number $W(a_1a_2\ldots
a_n)\equiv\sum_i(a_i-3)$ of a border, which is the number of
$60\degree$ turns undergone by a direction which is parallel
transported about the border.  The total net number of disclinations
within the border is $W+6$.  If we allowed seven-fold disclinations,
they would be subtracted from this number.  Since we only allow single
positive disclinations, we can conclude that the winding number around
any path on a valid capsid must be between $-6$ and $+6$, leaving $6-W$
disclinations which must be placed in the unfilled part (the other side
of the border, counting the vertices on the border itself).

%\subsection{Growth steps}
%We can express the growth steps in terms of their effect on the
%border.  Accretion takes $ab\to(a+1)1(b+1)$.  Joining takes
%$abc\to(a+c)$ where $b=5$ or 6.  Insertion takes $abc\to(a+1)(c+1)$
%where $b=4$ or 5.  Note that neither accretion, nor joining with
%$b=6$, nor insertion with $b=5$ changes the winding number.  We see
%from this that there exist invalid borders which are always
%incompletable.  For instance, $x555$.  Only insertion is allowed at
%the middle 5, leaving $x66$.  Now only joining is allowed for the 6 in
%the middle but if $x\ge 1$ then this leaves a vertex with more than
%six triangles.  [More on this?]

\subsection{Six disclinations remaining}
We will now show that any border with winding number $W=0$ which
does not intersect itself on a flat reference lattics is completable
by applying a finite number of growth operations to the border,
resulting in a self-complementary border of the form $3^m43^n2$, which
can be glued onto a copy of itself to make a complete capsid.

First draw the border on a flat reference lattice.  It is now clear
that triangles can be added to the border to transform it to the
required form.  So any capsid with a non-intersecting border and
$W\le0$ is completable.

\begin{figure}
  \includegraphics[width=\columnwidth]{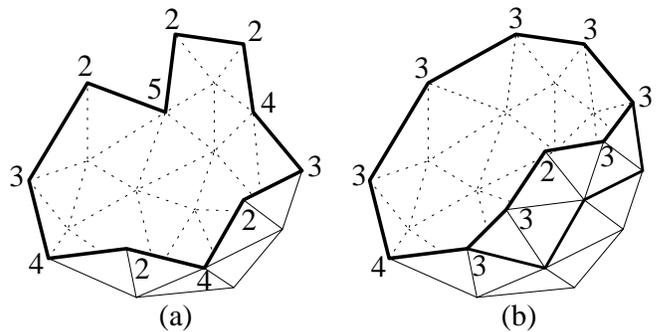}
  \caption{\label{fig:d6}Adding triangles to a $W=0$ border (a) to transform 
it into the self-complementary form $3^m23^n4$ seen in (b).}
\end{figure}

\begin{figure}
  \includegraphics[width=0.8\columnwidth]{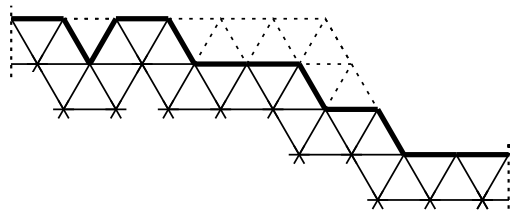}
  \caption{\label{fig:d6flat}An alternate point of view of the same
procedure as illustrated in FIG.~\ref{fig:d6}.  We flatten the border
onto a flat triangular reference lattice.  The dashed lines on the
left and right correspond to the place the border has been cut.  That
$W=0$ is evident because there is no net rotation after traversing the
border.  It is also clear that adding the dashed triangles results in
the same $3^m23^n4$ border as above.}
\end{figure}

This procedure is demonstrated in FIGS. \ref{fig:d6} and \ref{fig:d6flat}

\subsection{Late stage completability}
When $W>0$, there are more constraints.  We can no longer add
triangles freely since every row we add is smaller due to the enclosed
disclinations.  We will begin by considering the case of an incomplete
capsid with eleven disclinations enclosed, leaving a deficit of one
disclination needing to be placed.

\subsubsection{One disclination remaining}
\begin{figure}
  \includegraphics[width=\columnwidth]{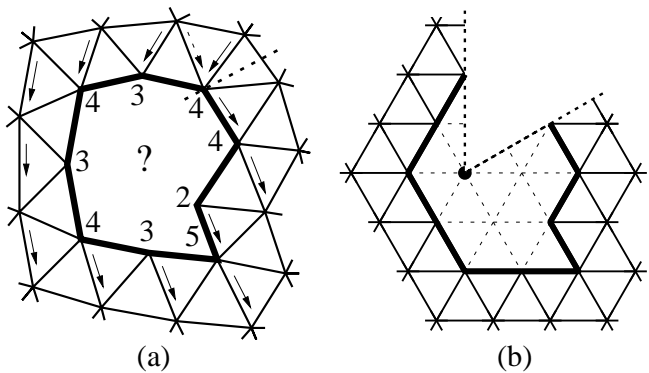}
  \caption{\label{fig:border} 
A border with $W=5$.  Parallel transporting
a direction around the border yields a rotation of $60\degree$.  A
disclination must therefore be located within this border.  The dashed
line shows where we plan to cut. (b) We see the same border flattened
onto a reference lattice.  The 60\degree\ rotation is now more clear.
The dashed lines are part of an equilateral triangle and therefore show
the required location of a disclination.  Any choice of cut results in
the same location, as long as the of the cut is chosen so that the triangle
is equilateral.}
\end{figure}

In this case we can easily look at the reverse picture.  If the border
is completable then it is a path on a valid complete capsid and we can
therefore look for a pseudo-complementary border to fill it.  We can
represent a triangular lattice with a single disclination as a flat
triangular lattice with a $60\degree$ section cut out and the edges
identified.  If we therefore flatten our border onto a flat lattice,
we expect the first and last points to be identified by this edge and
therefore we can draw an equilateral triangle with the third point at
the required location of the disclination.  While the edges of the
triangle need not be along a lattice direction, the third point is
necessarily on the lattice.  The border is completable if and only if
this disclination is at an unoccupied point (outside of the original
border).  Note that because the border has a $60\degree$
rotation, this point is unique, regardless of the choice of starting
and ending point. \SH{more here?}  This process is demonstrated in
FIG.~\ref{fig:border}.

\subsubsection{Two disclinations remaining}
\begin{figure}
\includegraphics[width=\columnwidth]{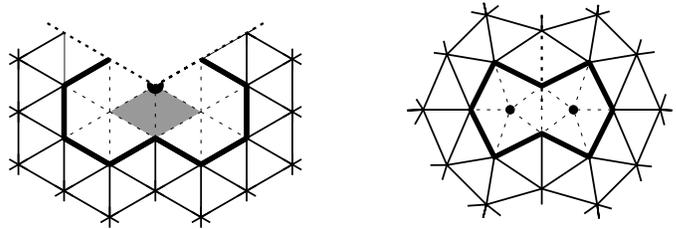}
\caption{\label{fig:disc2b}Rearrangement of a +2 ($120\degree$)
disclination located on a vertex into a pair of single disclinations
with the same border.  The two shaded triangles are removed.}
\end{figure}

\begin{figure}
\includegraphics[width=\columnwidth]{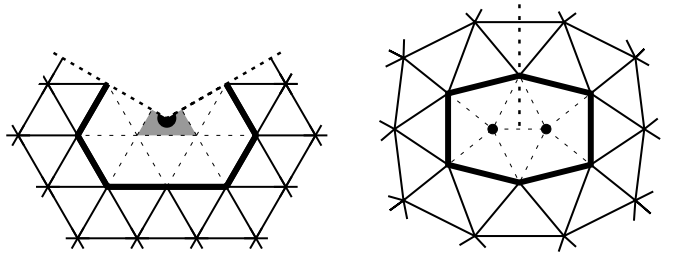}
\caption{\label{fig:disc2a}Rearrangement of a +2 ($120\degree$)
disclination located on a triangle into a pair of single disclinations
with the same border.  The shaded part of the triangle is removed.}
\end{figure}

Two disclinations ($W=4$) works in a very similar way to the single
disclination discussed above, except we have a
$120\degree$-$30\degree$-$30\degree$ isosceles triangle instead.  This
gives a single charge $+2$ disclination, but since we do not allow two
disclinations at the same point, we must move them slightly.
FIGS.~\ref{fig:disc2b}-\ref{fig:disc2a} show the two possible
situations and equivalent fillings with only single disclinations and
the same border.  If the center is on a lattice point, then the
disclinations can each move in opposite directions to neighboring
points and the same region of the plane will be cut out, up to a
triangle at the apex, as shown in FIG.~\ref{fig:disc2b}.  If the
center is in the center of a triangle rather than on a lattice point,
we can place the two disclinations on adjacent lattice points around
the triangle for the same effect, as shown in FIG.~\ref{fig:disc2a}.
The disclinations can be further separated in a similar fashion.
%After this, we can use normal disclination movement rules
%\todo{[disclination movement rules are not discussed in the paper?]}
%to move pairs around within the border.

This breaks down if the $+2$ disclination is on a vertex on the border
which has 4 or more triangles.  In this case there is no way to
separate the disclinations without one of them crossing the border.

\subsubsection{Three disclinations remaining}
\begin{figure}
\includegraphics[width=\columnwidth]{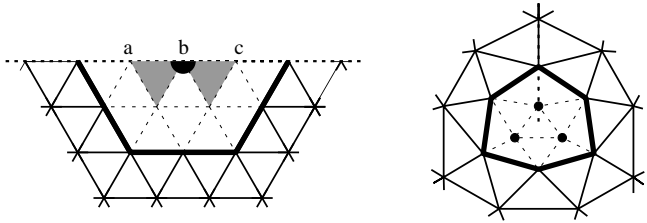}
\caption{\label{fig:disc3b}Rearrangement of a +3 ($180\degree$)
disclination located on a vertex into three single disclinations
with the same border.  The two shaded triangles are removed.
Note that vertices $a$, $b$, and $c$ all come together to form
a single five-fold vertex.}
\end{figure}

\begin{figure}
\includegraphics[width=\columnwidth]{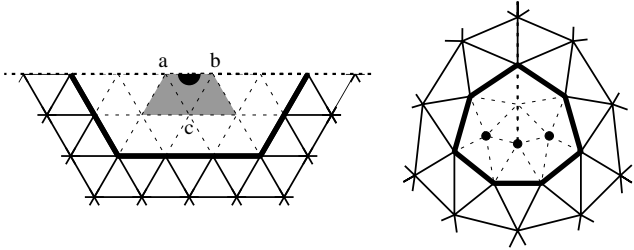}
\caption{\label{fig:disc3a}Rearrangement of a +3 ($180\degree$)
disclination located on an edge into three single disclinations
with the same border.  The three shaded triangles are removed and
the vertices $a$, $b$, and $c$ collapse to a single five-fold
vertex.}
\end{figure}

The case of $W=3$ follows the same way, except now we find a $+3$
disclination on the midpoint of a line segment joining the two
identified points.  This $+3$ disclination may be on a lattice point
or on the edge of a triangle.  Both can again be split similarly to
the previous case, as seen in FIGS. \ref{fig:disc3b}-\ref{fig:disc3a}.
As seen in the flattened pictures, the $+3$ disclination is always
within the border, provided the flattened border does not intersect
itself or the ``cut line''.  Thus outside
of these cases, the border is only incompletable if the $+3$
disclination cannot be split properly without any single disclinations
crossing a border.

\section{Steric considerations}
\label{app:steric}
Our triangular units are two-dimensional objects but they
represent three-dimensional structures in space. 
Thus, we must explicitly ensure that two triangles can never be
in positions, such that the proteins they represent would overlap in
space.  This appendix collects details concerning the implementation
of steric constraints.  First (Sec.~\ref{sec:steric-pot}) we write the
explicit form of the term in our Hamiltonian that prevents 
self-intersection;  then (Sec.~\ref{sec:steric-growth}) we discuss
the way in which steric constraints tend to assist growth and to 
discourage the wrong steps that lead to failure.

\subsection{Steric potential}
\label{sec:steric-pot}

%--- snip
The final term in eq.~(\ref{eq:ham}) was a steric repulsion term:
since our capsid units are two-dimensional triangles,
some such term has to be added by hand, to account for the thickness of our 
three-dimensional proteins and disfavor unphysical configurations.
The details of this term were deferred from Sec.~\ref{sec:model-ham}
to this appendix.
The steric term should have the simplest possible form,
in keeping with the toy-model spirit of our other terms.

In the steric term, the two kinds of degrees of freedom 
-- topological and positional -- clash in a sense.
Two units that are nearby in space may be many steps apart on the bond network,
and thus practically decoupled from each other.
(There is little interaction in the elastic energy, and
furthermore the ways they constrain the available discrete growth steps
are independent.) Hence, $\ham\steric$ must consist of 
topologically long-range, but positionally short-range, interactions.  

We chose an implementation based on augmenting each triangle by
another vertex over the face (on the interior side), 
thus forming a tetrahedron.
%%% pointing toward the center of the capsid. 
We define a repulsion between the apex vertex 
of each tetrahedron and every (non-apex) vertex
of every other triangle.  Thus,
\begin{equation}
  \ham\steric = \sum_{I,j} V\steric(|\vec r_I^\Delta-\vec r_j|),
  \label{eq:ham-steric}
\end{equation}
where $\sum_I$ is a sum over triangles and $\vec r_I^\Delta$ is an
equal distance $\ell\steric\lesssim r_0$ inward from the three
vertices of the triangle.  Furthermore, we require $V\steric(r)=0$ if
$r\ge\ell\steric$, which is the case for all pairs $I,j$ in most
capsids.  This form allows the edges of unconnected triangles to be
incident while maintaining $\ham\steric=0$ so long as the triangles do
not actually intersect.  

We choose the simplest form which is
differentiable at $r=0$ and $r=\ell\steric$,
\begin{equation}
  V\steric(r) = k\steric(\ell\steric^2 - r^2)^2,\quad r<\ell\steric,
\end{equation}
Choosing $\ell\steric\approx0.65r_0$ generally provides sufficient
stericity while not interfering with the shape of
non-self-intersecting capsids.

It is important to stress that this steric term should not affect
most capsids.  For non-growing capsids, we generally turn it off to
increase efficiency, since it always vanishes.

%\SH{I could increase $\ell\steric$ if
%I remove nearest-neighbor interactions here.}  
%\SH{I could also
%potentially add a topological constraint on the growth by checking
%each possible growth step against self-intersection on a flat lattice,
%but maybe that's too complex ($O(b^3)$) anyway} \SH{It occurs to me
%that this business of finding a suitable heuristic here is similar to
%finding growth rules which work at producing symmetric capsids.  In
%the latter case, there were times in the same growth where the
%threshold angle was both too large and too small.  Similarly we still
%get cases where the heuristic is both too restrictive (doesn't allow
%valid growth, distorts good capsids) and too lax (allows invalid
%growth and overlap).  What is the correct way to deal with this?}
%--- snip

%\todo{``It is important to stress that this steric term should not affect
%most capsids... ''
%``For non-growing capsids, we can even probably turn it off''
%Please delete/include/convert into comments.}

\subsection{Steric growth heuristics}
\label{sec:steric-growth}

%\SH{steric heuristic - necessary for triangle model ``An additional drawback...
%''}
%First, we deal with the steric constraints by prohibiting accretions
%and insertions which would cause the new triangle to have a nonzero
%steric potential with any other vertex [only vertices around the
%border are tested?], or cause the centroids to be too close (within
%$\ell\steric/\sqrt{10}$ or cause any intersection of edged when viewed
%from a point along the new triangle's norm from the midpoint of the
%edge.  We also prohibit any moves which would cause anything other
%than a pentamer or hexamer to be enclosed.

%Regarding the energy parameters, we can ignore $r_0$, since it merely
%sets a length scale for our capsids.  Further we can largely ignore
%$k\steric$ and $\ell\steric$ because they are mostly irrelevant beyond
%their role of preventing self-intersection.  

While the steric potential discussed in Sec. \ref{sec:steric-pot} is
useful to prevent capsids from relaxing to unphysical positions, it does
not directly help the growth rules.  Because growth rules are based
entirely on rates $k_A$, $k_I$, and $k_J$ derived from the local geometry
around individual vertices, there is no way to directly determine whether
a step will cause a steric hindrance.  Because such growth steps are
not likely to occur in nature, we implement a heuristic to detect such
steps and remove them from the set of allowed growth steps by setting
the rate to zero.

Before any accretion or insertion, we perform two tests.  First we
look at the steric potential $\ham\steric$.  If the accretion causes
$\ham\steric\ne 0$ then the accretion fails.  Next, if the accretion
causes the centroid of one triangle to be within
$\ell\steric/\sqrt{10}$ of the vertex of another triangle, then the
accretion fails.  This is necessary because the first test misses the
case where two triangles are directly on top of one another.  This
case is less important while minimizing, because minimization would
need to pass a large energy barrier, while growth steps can jump over it
for free.

\bibliography{capsid,footnote}
%\bibliography{experiment}
\bibliographystyle{apsrev}

\end{document}